# DEEP BRAIN STIMULATION FOR PARKINSON'S DISEASE:

## A SURVEY OF EXPERIENCES PERCEIVED BY RECIPIENTS AND CARERS

by

N.W. Page[1,2], C. Hall[2] and S.D. Page[2]

Revision 1

August 2015

[1] Corresponding author, The University of Newcastle, Australia     eml neil.page@newcastle.edu.au

[2] DBS Support Group of Parkinson's Queensland Inc.

# Contents

# Figures

# Tables

# Acknowledgements

This project was an initiative of the *Deep Brain Stimulation Support Group of Parkinson's Queensland.* It would not have been possible without the willing participation of people who have had Deep Brain Stimulation for Parkinson's disease, their carers and families. For reasons of confidentiality they cannot be named but they know who they are. To them we give heartfelt thanks and good wishes for their continuing journey with Parkinson's disease. Your contributions to this study will be of great help to those that follow in your footsteps.

Our thanks also go to those who read drafts of this report and provided comments and input to help improve it.  John Quinn and Em Prof Ludvik Bass were especially helpful in this regard.

This was an independent project resourced entirely by the DBS Support Group of PQI. No benefits were gained by any participants or contributors to the study.

While every effort has been made to ensure the accuracy of information contained in this report, those considering DBS for Parkinson's disease should base their decisions on advice from suitably qualified medical practitioners. PQI, its staff and volunteers disclaim liability for any loss or damage that may arise as a result of any person relying on information given in this report.

# Preface

The decision to have the Deep Brain Stimulation (DBS) procedure is often a daunting one for many of those with Parkinson's disease (PD) and who have the option of receiving this treatment. It can also be daunting for carers. Many find the decision to have elective brain surgery a difficult one, a decision often made with limited knowledge or understanding of possible outcomes.

Parkinson's disease is a complex neurological disorder with a wide range of motor and non-motor symptoms. Those living with the disease often seem to have an almost unique combination of its symptoms. People contemplating DBS are naturally curious about the outcomes they may experience. Their carers, family and friends share in this curiosity.

It is widely recognised that successful outcomes from DBS for PD depend in part on those involved having realistic expectations of the outcomes. Those involved include the person with PD, their carer, family and friends. Information about DBS is available from the clinical team responsible for the procedure. More is available from published medical literature, but this isn't always in a form that the general public have access to, or are able to fully understand.

Those that have already had DBS for PD are another valuable source of information. The challenge with this source is to ensure that the information gained in this way is representative. The feedback we receive from our support group network, in particular from the *Deep Brain Stimulation Support Group of Parkinson's Queensland,* suggested that the response to DBS is highly variable between individuals and can also vary over time. Further, perceptions of outcomes from the DBS procedure can sometimes be different for DBS recipients and carers. Recognition of this possible range of outcomes was the catalyst for the survey described here, the goal being to provide broadly based information about outcomes collected in a systematic way for a representative number of cases.

The results from this study are intended to better inform those contemplating the procedure about the range of outcomes and how these outcomes can change over time, based on experiences reported by those who have had DBS. It is also directed to those who have had the procedure and are interested to learn how their experiences compare with others. Carers, families and friends may also find useful information from this survey better equipping them to support DBS recipients during the post-DBS journey. It may also help service planners and providers identify areas of need guiding the development of more targeted services. As such it complements the efforts of organisations such as the Cochrane Collaboration (http://www.cochrane.org/) devoted to providing accessible systematic reviews of treatment outcomes to provide additional guidance for health decisions.

<u>Revision 1</u>

Further analysis has been conducted on the information collected in this survey, guided by feedback from readers. As a result, this revision includes more information about the DBS experience profile of participants and some disaggregation of scores from DBS recipients and carers showing more clearly any differences in assessments of DBS outcomes from these two groups. As a consequence the discussion has been largely rewritten and restructured to improve clarity. Some minor errors have also been corrected. The authors would like to thank readers for their helpful comments.

NWP

August 2015

# GLOSSARY

| | |
|---|---|
| Bradykinesia | Slowness of movement |
| C | Carers |
| DBS | Deep brain stimulation |
| DDS | Dopamine dysregulation syndrome, characterised by self-control problems |
| DRT | Dopamine replacement therapy |
| Dyskinesia | Random jerky movements, a side effect of high levels of levodopa intake |
| GPi | Globus pallidus internus |
| ICD | Impulse control disorder, characterised by failure to resist a harmful urge or impulse |
| PD | Parkinson's disease |
| PPN | Pendunculopontine nucleus |
| PQI | Parkinson's Queensland Incorporated |
| R | Recipients |
| STN | Subthalamic nucleus |
| VIM | Ventralis intermedius nucleus of thalamus |

# SUMMARY

This study was aimed at providing additional insight into the outcomes from DBS for Parkinson's disease, how these vary between people who have had the procedure, and over time.

New perspectives have been gained on the outcomes from DBS for Parkinson's disease by sampling the experiences of both recipients of DBS and carers. Data were collected mostly by structured interview from 52 cases involving 91 participants – 46 recipients and 45 carers. Post-DBS experience ranged from 10-129 (av. 40.1) months at the time of interview.

There were significant variations in perceived outcomes over time. Some experienced extreme variations as a consequence of episodes involving hardware and other problems requiring additional surgery. This was reported in 27% of cases. Despite these complications and adverse effects, most experiencing these episodic difficulties went on to ultimately report good outcomes. More generally there were often noticeable changes in the outcomes from DBS with time – even within the first 12 months following the procedure, with some symptoms showing sustained improvement and some showing reducing benefits with time.

Holistic assessments of experiences following DBS for PD were largely positive, but in some cases there were noticeable differences in the assessments by recipients and carers. For assessments valid at the time of interview 26 recipients and 17 carers, together representing 58% of cases, commented that the outcome was good. A second group of 11 recipients and 12 carers, together representing 31% of cases, reported mixed results but overall a positive experience. A third group of 6 recipients and 8 carers, together representing 19%[1] of cases, reported negatively about the outcomes. Following DBS overall quality of life was considered, at the time of interview, to be better by28 recipients and 25 carers but worse by 6 recipients and 10 carers. Post-DBS experiences of both motor and non-motor symptoms varied greatly between cases. With all symptoms, some participants reported that they were worse following DBS, but in the great majority of cases symptoms were reported to be the same or better following DBS.

When considering the average of participant responses, tremor and dyskinesias were considered better or much better following DBS, with benefits sustained with time.

Twelve months after DBS many symptoms were on average considered the same or better after DBS, but for many, some decline in benefits was apparent over this period.

Some symptoms were reported to show no improvement, or be worse following DBS. Twelve months after the procedure the average of participant responses indicated that symptoms including speech, postural stability, swallowing, handwriting, cognitive function and incontinence were worse.

There was broad agreement between the assessments of recipients and carers for the outcomes for many symptoms, but some differences were noticeable. At some times after DBS, carers reported greater improvements than recipients for a range of symptoms including facial expression, difficulty walking, slowness or lack of movement, difficulty standing from a chair and freezing. In contrast, carers were more likely to give less positive assessments than recipients for symptoms including sleep quality, fatigue, cognitive function, mood and behaviour, anxiety, vision, depression and quality of life.

---

[1] Percentages don't total 100 because in some cases recipients and their carers gave different responses.

# INTRODUCTION

Deep brain stimulation for Parkinson's disease is now a well-established surgical approach to the management of Parkinson's disease[1]. It involves placing electrodes at key sites deep in the brain to provide continuous electrical stimulation to those sites. DBS can be applied to just one side of the brain (unilateral DBS) or both sides (bilateral DBS). The electrical power and signals used for stimulation are provided by a device surgically implanted under the skin, usually in the chest area, with electrical leads running under the skin to the electrodes in the brain.

The history of DBS dates back to at least the 1950's. In its earliest implementation it was used as a tool in the surgical treatment of psychiatric illnesses[2]. Its potential for the treatment of movement disorders, including PD, was demonstrated in 1987[3]. When applied to the subthalamic nuclei (STN) it was shown in 1993 to provide significant improvements to symptoms of PD[4] and in 1998 reduced need for dopamine replacement drugs[5].

Evidence of the benefits and side effects of the procedure are widely documented in the medical literature. Extensive reviews of this literature have been conducted by others[6,7,8] and further review of this medical nature is beyond the scope of this report. However the principal findings of these reviews will be summarised to provide background for the study reported in this document.

Control of motor symptoms of PD is the main treatment goal with DBS[9]. Over time a number of structures in the brain have been investigated as targets for DBS, shown in Table 1.

| Table 1- Sites for DBS Sometimes Used in the Treatment of PD[10] | | |
|---|---|---|
| Site | Acronym | Comments |
| Ventralis intermedius nucleus of thalamus | VIM | For tremor, but commonly STN and GPi are now more favoured because they can also improve tremor plus other PD symptoms. |
| Subthalamic nucleus | STN | Generally now considered the optimal choice for PD, depending on PD symptoms. |
| Globus pallidus internus | GPi | Sometimes preferred depending on symptoms, for example when treating slowness of movement (Bradykinesia) and rigidity. |
| Pedunculopontine nucleus | PPN | Sometimes preferred depending on symptoms, for example when treating gait disorders and postural instability (axial symptoms). Fasano (2012) reports outcomes for this site have been very variable and consequently interest in it seems to have waned. |

There are no guidelines for the choice of target for DBS. STN seems to offer the best overall management of PD symptoms, but there are special cases when other sites for DBS may be more advantageous.

The outcomes from DBS depend to some degree on the target site chosen. Dopamine replacement therapy is a good predictor of outcomes following STN DBS – if a symptom responds to levodopa medication there is a good chance that it will respond to STN DBS[11]. On average STN DBS leads to a reduced need for levodopa by about 55% with consequential improvements in dyskinesia. This reduced need for levodopa is not duplicated with GPi DBS[12].

There is good evidence to show that, for many, DBS improves many motor symptoms especially when combined with dopamine replacement therapy[13,14]. In the case of STN DBS for which long term data is available, improvements in motor symptoms and dyskinesias are sustained for at least 10 years, but improvements in slowness of movement (bradykinesia) wear off over time. DBS to STN, GPi and VIM targets can all relieve tremor. Gait and balance problems can be resistant to both dopamine replacement therapy (DRT) and DBS. DRT and STN DBS have about the same effect on postural instability and gait problems but both applied together can give short term improvements above the single treatment regime. Despite STN DBS, gait and balance problems get worse over time. Speech can be adversely affected by DBS. Speech worsens after STN DBS – 56% have worsening speech one year after implantation, rising to 90% at 8 years. At one year after implantation, loudness is increased but intelligibility decreased[15].

Non-motor symptoms of PD also respond to DBS[16]. In addition to the worsening speech already mentioned above there is a slight but significant decline in episodic memory, executive function and abstract reasoning, but this decline is similar to that observed in patients on drug therapy alone. Bilateral GPi DBS has fewer cognitive or behavioural adverse effects than STN DBS , at least over a 10 year period. Pre-existing impulse control disorders (ICDs) are markedly improved with STN DBS, but some studies have reported ICDs after DBS. Through its ability to reduce the need for DRT, STN DBS can improve dopamine dysregulation syndrome (DDS), a syndrome characterised by self-control problems such as gambling, hypersexuality or addiction to medication.

Apathy, mood disorders and anxiety can all be affected by DBS[17]. Apathy worsens with STN DBS. At about 5 months after surgery 54% suffer apathy which is reversible in half of recipients at 1 year. Mood disorders (depression or mania) can occur after STN DBS in either transient or persistent form. Following STN DBS depression can get better, stay the same or get worse. Suicidal tendencies have been reported following STN DBS – 0.9% of patients attempting and half of these being successful. Suicide rates are highest in the first year after the surgery. Manic symptoms occur in about 4% of those given STN DBS, sometimes in the immediate post-operative period. Manic symptoms can last for hours or a few days and seem to be linked to the DBS stimulation and sometimes respond to different stimulation settings. At 7 months after surgery there are no overt mood variations. Other DBS targets occasionally affect mood, but to a lesser extent that STN – some in fact leading to improvements. Various effects on anxiety have been reported for STN DBS – some reporting better, the same or worse anxiety following DBS.

Bilateral STN DBS improves continuous sleep time, sleep efficiency and decreases nocturnal dystonia. These improvements relate to improvement in bradykinesia, increase in bladder capacity and reduced daytime sleepiness because of reduced DRT. There is some evidence that STN DBS leads to improvement in restless leg syndrome, but other studies indicate that it is made worse by the reduction in DRT following STN DBS. PPN DBS can significantly increase REM sleep, but there is little evidence that other DBS sites influence sleep in this way.

Like any medical procedure, DBS can lead to complications and unwanted side effects (adverse effects). The incidence of these outcomes varies significantly between centres performing the DBS procedure[18,19]. In the case of STN-DBS, published results are summarised in Table 2.

Few reviews have been published that have included subjective assessments by patients or carers on their perceptions of outcomes. A clinical review of patients complaining of "failed" DBS procedures[20] revealed a wide range of circumstances influencing their opinions. These included errors in original diagnosis, misplaced leads and hardware complications and DBS programming inadequacies. With appropriate intervention about half these patients ultimately had good outcomes. In another study[21] the expectations and subjective perceived outcomes of 30 patients were assessed before and 3 months after surgery. In terms of perceived subjective outcomes, 8 (26.7%) were negative, 8 (26.7%) were mixed and 14 (46.7%) were positive. Those reporting negative outcomes were characterised by unrealistic expectations and higher apathy and depression scores, before and after surgery, compared with those reporting more positive outcomes. In another study differences in perceptions about quality of life aspects were reported for DBS recipients and carers[22]. While quality of life was significantly improved by DBS, at 12 months following the procedure carers ratings of quality of life were lower than those given by recipients.

| Table 2– Some Complications and Adverse Effects after STN-DBS[23,24] | |
|---|---|
| Complication | Incidence (range reported from different DBS performing centres) |
| Haemorrhage | 0.2 – 12.5% |
| Transient post-operative confusion | 1 – 36% |
| Infection | <1 – 15.2% |
| Electrode breakage | 0 – 15.2% |
| Electrode migration or misplacement | 0 – 18.8% |
| All hardware related complications (skin erosions, wire breakages, battery or electronic malfunction) | 2.7 – 50% |

The present study was designed to provide further information about outcomes from DBS for Parkinson's disease from the point of view of both those who have had the procedure and their carers. Throughout this work the word "carer" is used loosely to describe anyone who has a close personal knowledge of the condition of the person with PD before and after DBS surgery. It might be a spouse, partner, adult child or friend. The study goal was to provide additional information about the range of perceived outcomes and how these vary over time to help patients and their loved ones form more realistic expectations of the procedure, and through that, contributing to less stressful and better outcomes.

The results of this study should be interpreted with caution. Parkinson's disease is progressive and DBS is not a cure. This study was limited to people who have had DBS for PD. Comparison with others who were being treated non-surgically was beyond the scope of this work but has been reported by others[25,26]. Also a number of DBS recipients participating in this study had other health problems which may, in ways difficult to quantify, have influenced their perception of the progress of Parkinson's disease.

## METHODS

### Participants

Participants were drawn from the DBS support network of Parkinson's Queensland. An important decision made in scoping this survey was that there would be no exclusions – all outcomes were valid outcomes. In a few cases this meant that responses could only be obtained from carers because of physical or mental impairment of recipients.

Participants (either recipients of DBS, carers or both) for about 50 cases were sought to provide a significant sample of perceived outcomes. As it transpired, progressive analysis of the responses obtained suggested that established patterns of outcomes were not changing significantly at about this number (saturation).

Those invited to participate were chosen in date order of the DBS procedure, (known for those having the procedure 2011 onwards), otherwise in alphabetical order based on surname. Invitations included a written copy of the structured questionnaire that was the survey instrument used in this study thus allowing invited participants to make an informed decision as to whether or not they wished to be included. Of those approached to be involved in the study six recipients and their carers declined to participate and a further 25 did not respond to the invitation.

This study was directed towards cases involving post-DBS experience - of about 12 months or longer. Agreement to participate was based on pre-view of the survey instrument. The study was guided by the values and principles of ethical conduct for human research specified by Australian authorities[27].

## Procedure

The survey was based on structured interviews, where possible with both DBS recipients and their carers. The questions for the interviews were workshopped by members of the DBS support group of PQI. They reflect those issues seen as important by members of that group, recognising that post-DBS experiences for many vary over time. Details of the survey instrument are shown in Appendix 1. There were essentially 4 parts to the survey instrument: (1) a preamble describing the background, purpose, scope and confidentiality undertakings, (2) questions relating to personal background of the DBS recipient and procedure history, (3) PD symptoms following DBS – Part A, important changes noticed by both DBS recipient and the carer, in their own words, and (4) PD symptoms following DBS – Part B, a subjective scaling of changes in PD symptoms by both DBS recipient and carer. Responses for (3) and (4) were sought over several time intervals covering the first 12 months following DBS, nominally 1, 6 and 12 months after the procedure and also at the date of interview.

The special focus on the first 12 months after DBS was determined by frequent anecdotes from our members about significant changes over this "settling in" period during which time there are often significant adjustments to the DBS settings as part of standard clinical practice.

Given that participants were widely scattered geographically, nearly all interviews were conducted by telephone. DBS recipients and their carers were asked whether they would like to be interviewed together or separately. Most chose to be interviewed together. Whether interviewed separately or together there were frequently differences in responses to some questions.

Responses were transcribed and recorded on a *case* basis (i.e. recipient and/or carer responses relating to a particular procedure) with personal identities removed. Responses for Part 3 of the survey instrument (PD symptoms following DBS – Part A) were analysed using a form of content analysis and the graded responses for Part 4 (PD symptoms following DBS – Part B) reported graphically to more clearly demonstrate the variations in time.

## RESULTS & DISCUSSION

### Background

Of the 52 cases studied, the gender balance for those who had the DBS procedure was 27 males and 25 females with birth years ranged from 1933 to 1968. Interviews were conducted over 2012 and 2013. At the time of interviews, recipients' post-DBS experience ranged from 10 to 129 months (Fig. 1) with the average 40.1 months.

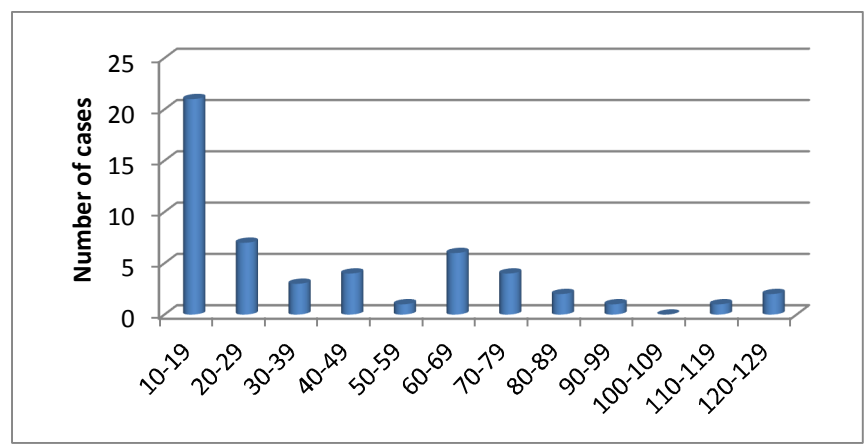

**Fig. 1  Months following DBS procedure for participating recipients**

In all but 2 cases (10 and 11 months) post-DBS experience was 12 months or longer.  In 40 cases interviews were possible with both the DBS recipient and the carer, in 6 cases with the recipient only and in 5 cases with the carer only so that in all there were 91 participants.  In those cases in which the recipient only was interviewed, a carer wasn't available or declined to participate. In the cases for which a carer only was interviewed, the DBS recipient could not communicate effectively because of impairment in writing, speech ability or cognitive function. In one case in which the DBS recipient and carer were interviewed, Part B of the structured interview had to be abandoned because of the emotional state of the carer and the cognitive impairment of the recipient.

Most participants did not know the location in the brain that was the target for DBS (Table 1). Except for one known case involving PPN DBS, the few who did know identified the subthalamic nucleus (STN) as the target in their case. Because of this uncertainty, no analysis of outcomes to DBS could be performed on the basis of DBS target. Also, while most participants recalled that there was a reduction in PD medication post-DBS, most could not recall with accuracy the details of their medication for PD over the time period covered by the survey. Any further attempt at analysis of this aspect was therefore abandoned.

Fourteen of the 52 cases (27%) reported further surgery related to DBS not including routine and expected battery replacement. Some had multiple procedures – this group of 14 reported at least 22 surgical procedures. Reasons for these procedures included relocation of leads or probes (5), replacement of broken leads (5), replacement of a faulty stimulator (2), hardware eruption through skin (1 – multiple sites), relocation of stimulator (2) and infection (4). In those cases involving infections, all or some of the implant was replaced. The timing of these post-DBS procedures ranged from 1 to 72 months following the initial implantation. These reported complications are consistent with published statistics on these events summarised in Table 2.

## Part A – Most Important Changes Noticed by Participants Following DBS Procedure

The spontaneous responses of participants touched on over 100 topics. These were sorted and, where closely related, aggregated as indicated in the following tables. Caution must be exercised in interpreting some of the responses reported here and in subsequent sections because a number of participants had other health problems in addition to Parkinson's disease (comorbidities). Some could be considered secondary to PD (e.g. fractures caused by falls, and their complications) while others were quite separate illnesses (e.g. cancer, arthritis). In cases where comorbidities existed, participants were asked to differentiate as well as they could the separate health impact from PD, the focus of this study.

Participant responses to Part A have been grouped on a category basis and are presented below in order of mention frequency at the time of interview, the order thus being in terms of longer term effects. For several reasons – different responses from DBS recipient and carer, or the lack of contribution on a particular topic by either recipient or carer – the number of **cases** represented by the tabulated data is usually less than the sum of recipient and carer counts.

### 1. General Comments on DBS Experience

Most participants offered holistic reflections of their DBS experience. These are summarised in Table 3 in which similar comments are clustered. Some who commented that they would or would not agree to the procedure again are included in other comment counts (rows) of Table 3.

Positive comments about the procedure varied in number over the first 12 months after the operation (Table 3), with more frequent mentions in the early months. Comments reporting a good outcome, at least at some time following DBS, were made in 41 of the 52 cases (79%), by either the recipient or carer or both.

The longer term assessments were those reported at the time of interview. At that time a large proportion of participants still reported a good outcome, in several cases despite transient difficulties in the post-DBS period. Those that mentioned good outcomes at this time included 26

DBS recipients, and 17 Carers. Considered on a *case* basis a good outcome was reported by either the recipient or carer or both in 30 cases of the 52 in the study (58%).

Another group reported mixed outcomes, but on balance considered DBS a positive experience. At the time of interview this group included 11 recipients and 12 carers, collectively representing 16 of the 52 cases (31%).

A third group reported negative experiences at various times following DBS. At least at some period following DBS negative experiences were reported in 13 cases (25%) by either the recipient or carer or both. However the longer term assessments in this group reported at the time of interview comprised 6 recipients and 8 carers representing 10 of the 52 cases (19%).

| Table 3– Numbers of General Comments on DBS Experiences | | | | | | | | | | |
|---|---|---|---|---|---|---|---|---|---|---|
| | DBS recipient | | | | Carer | | | | Cases total | Cases at Int |
| **Months after initial DBS procedure** | 1 | 6 | 12 | Int | 1 | 6 | 12 | Int | | |
| **Comments** | | | | | | | | | | |
| Felt a lot better, things going well, better quality of life, good outcome | 11 | 4 | 7 | 26[a] | 11 | 10 | 7 | 17[a] | 41 | 30 |
| Mixed results but on balance positive experience | | | | 11[b] | 1 | 2 | | 12[b] | 16 | 16 |
| Bad experience, didn't work, expectations not met, DBS switched off or removed | 2 | 1 | 3 | 6[c] | 1 | 1 | 1 | 8[d] | 13 | 10 |
| Would do it again | | | | 4[e] | | | | 2[f] | 5 | 5 |
| Never do it again, wished it had never been done | | | | 2[g] | | | | 1[h] | 3 | 3 |

At interview post-DBS experience ranged from [a] 12-122, [b] 12-73, [c] 12-74, [d] 14-74, [e] 14-74, [f] 74-84, [g] 12-74, [h] 20 months

It is important to note that the sum of cases for good, positive and negative responses don't total to 52 because some participants views changed over the post-DBS period and some DBS recipients had different views to carers.

## 2. Comments Related to Motor Symptoms

Many of the comments made in Part A of the survey related to changes in patient symptoms following the procedure. Those comments relating to motor symptoms are grouped here in rank order of the number of cases where the comments were made valid for the time of interview.

Posture and Balance

Frequent comments were made about posture and balance. These are shown in Table 4 where related terms have been clustered. A number of DBS recipients and carers reported balance problems, falls or both with increasing frequency over the first 12 months following the procedure. At some time post-DBS these problems were reported by the recipient, carer or both in 20 of the 52 cases (38%). Reports relating to the longer term presence of these problems were those relevant at the time of interview (up to 112 months for recipients and up to 66 months for carers). These indicated balance and falls were a problem in 12 of the 52 cases (23%) at the time of interview.

In contrast, some participants reported better balance and fewer falls. Reports of this nature were received from the recipient, carer or both at least at some time following DBS in 8 of the 52 cases (15%) and at the time of interview in 2 of the 52 cases (4%).

| Table 4– Number of Comments About Posture and Balance | | | | | | | | | | |
|---|---|---|---|---|---|---|---|---|---|---|
| | DBS recipient | | | | Carer | | | | Cases total | Cases at Int |
| **Months after initial DBS procedure** | 1 | 6 | 12 | Int | 1 | 6 | 12 | Int | | |
| **Comments** | | | | | | | | | | |
| Balance problems, having falls | 4 | 7 | 12 | 9[a] | 3 | 6 | 9 | 7[b] | 20 | 12 |
| Better balance, fewer falls | 2 | 1 | 1 | 1[c] | 3 | 2 | 2 | 1[d] | 8 | 2 |
| Posture getting worse | | | | 1[e] | | | | | 1 | 1 |

At interview post-DBS experience ranged from [a] 5-112, [b] 14-66, [c] 27, [d] 122, [e] 12 months

Speech

Reports relating to speech indicated that for many, speech was perceived to worsen following DBS. Speech difficulties were mentioned by the recipient, carer or both at some time post-DBS in 21 of 52 cases (40%) and valid at the time of interview, in 13 of 52 cases (25%). In contrast, improved speech was mentioned in a couple of cases.

| Table 5 – Number of Comments About Speech | | | | | | | | | | |
|---|---|---|---|---|---|---|---|---|---|---|
| | DBS recipient | | | | Carer | | | | Cases total | Cases at Int |
| **Months after initial DBS procedure** | 1 | 6 | 12 | Int | 1 | 6 | 12 | Int | | |
| **Comments** | | | | | | | | | | |
| Speech worse | 2 | 2 | 5 | 8[a] | 5 | 6 | 1 | 7[b] | 21 | 13 |
| Speech better | | | 1 | 1[c] | | | 1 | | 2 | 1 |

At interview post-DBS experience ranged from [a] 12-83, [b] 24-83, [c] 60 months

Gait and Movement

Reports of movement and gait problems were also common. These were mentioned at some time post-DBS in 16 of 52 cases (31%) and at the time of interview, in 6 of 52 cases (12%). In contrast, better walking was reported in 13 of 52 cases (25%) at some time during the post-DBS period and in one case (2%) at the time of interview. Of note was the relatively high reporting, by both recipient and carer, of improved walking about one month after the procedure but this reporting was not sustained over the longer periods.

Other changes in movement or mobility were mentioned in a small number of cases (Table 6).

| Table 6 – Number of Comments About Gait and Other Movement | | | | | | | | | | |
|---|---|---|---|---|---|---|---|---|---|---|
| | DBS recipient | | | | Carer | | | | Cases total | Cases at Int |
| **Months after initial DBS procedure** | 1 | 6 | 12 | Int | 1 | 6 | 12 | Int | | |
| **Comments** | | | | | | | | | | |
| Slower movement, freezing, gait problems | 2 | 6 | 6 | 1[a] | 1 | 2 | 2 | 5[b] | 16 | 6 |
| Walking better | 6 | 1 | 1 | | 6 | 2 | 2 | 1[c] | 13 | 1 |
| Handwriting worse | | 2 | 1 | 1[d] | 1 | | | | 3 | 1 |
| Freezing improved, eliminated | | | | 1[d] | | | | | 1 | 1 |
| Unable to walk | | | | 1[e] | | | | | 1 | 1 |
| Handwriting better | | | | | 1 | | | | 1 | 0 |
| Spasms in legs | | | 1 | | | | | | 1 | 0 |

At interview post-DBS experience ranged from [a] 37, [b] 11-83, [c] 12, [d] 14, [e] 74, months

Tremor

Improvements in tremor were reported, particularly in the period immediately following the procedure (Table 7). At some time in the post-DBS period improvements in tremor were reported in 16 of 52 cases (31%) and at the time of interview in 6 of the 52 cases (12%).

In some cases tremor was reported to have returned. This was mentioned at some time in the post-DBS period in 5 of 52 cases (10%) and at the time of interview in 3 of 52 cases (6%).

| Table 7 - Number of Comments About Tremor | | | | | | | | | | |
|---|---|---|---|---|---|---|---|---|---|---|
| | DBS recipient | | | | Carer | | | | Cases total | Cases at Int |
| Months after initial DBS procedure | 1 | 6 | 12 | Int | 1 | 6 | 12 | Int | | |
| Comments | | | | | | | | | | |
| Tremor improved | 12 | 4 | | 6[a] | 4 | 2 | 2 | 1[b] | 16 | 6 |
| Tremor back | | 2 | | 2[c] | 1 | | | 1[d] | 5 | 3 |

At interview post-DBS experience ranged from [a] 14-97, [b] 31, [c] 11-27, [d] 66 months

Stiffness, Rigidity and Dystonia

There were a small number of reports relating to stiffness, rigidity or dystonia (Table 8). Reduced stiffness or rigidity was reported at some time in the post-DBS period in 6 of 52 cases (12%) and at the time of interview in 1 case (2%). Reports of improvements in dystonia at some time in the post-DBS period came from 3 in 52 cases (6%). Other features in this grouping were mentioned infrequently, Table 8.

| Table 8 - Number of Comments About Stiffness, Rigidity and Dystonia | | | | | | | | | | |
|---|---|---|---|---|---|---|---|---|---|---|
| | DBS recipient | | | | Carer | | | | Cases total | Cases at Int |
| Months after initial DBS procedure | 1 | 6 | 12 | Int | 1 | 6 | 12 | Int | | |
| Comments | | | | | | | | | | |
| No or reduced rigidity, stiffness | | 3 | | 1[a] | 2 | | | | 6 | 1 |
| Stiff neck | | | | 1[b] | | | | | 1 | 1 |
| Dystonia improved | 1 | | | | 2 | | | | 3 | 0 |
| Stiffness worse | 1 | | 1 | | | | | | 2 | 0 |
| Slowness of movement | | | 1 | | | | | | 1 | 0 |
| Arm paralysis | 1 | | | | | | | | 1 | 0 |
| Dystonia worse | | | | | | | 1 | | 1 | 0 |

At interview post-DBS experience ranged from [a] 97, [b] 17 months

Other Motor Symptoms

Reports of changes in other motor symptoms were relatively small in number, Table 9. In the immediate period following DBS, carers reported a loss of the characteristic masked (Parkinson's) face in 7 of 52 cases (13%). In some cases this masked face returned, at the time of interview in 2 of 52 cases (4%).

| Table 9 - Number of Comments About Other Motor Symptoms | | | | | | | | | | |
|---|---|---|---|---|---|---|---|---|---|---|
| | DBS recipient | | | | Carer | | | | Cases total | Cases at Int |
| Months after initial DBS procedure | 1 | 6 | 12 | Int | 1 | 6 | 12 | Int | | |
| Comments | | | | | | | | | | |
| "Parkinson's face" returned | | | | 1[a] | | 1 | | 1[b] | 2 | 2 |

| | | | | | | | | | |
|---|---|---|---|---|---|---|---|---|---|
| Difficulty swallowing, drooling | | | | 2[c] | | 1 | | 2 | 2 |
| Manipulation skills worse | 1 | | 1 | 2[b] | | 1 | 1[b] | 2 | 2 |
| Involuntary eyelid closure (blepharospasm) | | | 1 | 1[d] | 1 | 1 | | 3 | 1 |
| Restless legs | | | 1 | 1[e] | | | | 1 | 1 |
| No more restless legs | | | | 1[f] | | | | 1 | 1 |
| "Parkinson's face" gone | | | | | 6 | 2 | | 7 | 0 |
| Weak legs | 1 | | 1 | | | | | 1 | 0 |
| Manipulation skills better | | 1 | | | | | | 1 | 0 |

At interview post-DBS experience ranged from [a] 17, [b] 12, [c] 27-37, [d] 34, [e] 23, [f] 84 months

### 3. Comments Related to Non-Motor Symptoms

Non-motor symptoms also generated a significant number of comments. These follow, once again in order of the number of reports describing views at the time of interview.

<u>Memory and Judgment</u>

Reports of the kind "impaired judgement, impaired cognitive function or memory loss" were more common from carers, Table 10. In all, reports of this nature were received from the recipient, carer or both in 15 of 52 cases (29%) at some time post-DBS and for 9 of 52 cases (17%) at the time of interview. The higher count over the entire post-DBS period, in contrast to comments valid at the time of interview is consistent with some of the reports of this nature being generated by episodic events occurring in the first 12 months following DBS. These included short term disorientation, poor judgement while driving, examples of transient poor financial decision making and atypical or irrational shopping.

| Table 10- Number of Comments About Memory, Judgement and Intellectual Function | | | | | | | | | Cases total | Cases at Int |
|---|---|---|---|---|---|---|---|---|---|---|
| | DBS recipient | | | | Carer | | | | | |
| Months after initial DBS procedure | 1 | 6 | 12 | Int | 1 | 6 | 12 | Int | | |
| **Comments** | | | | | | | | | | |
| Impaired judgement, cognitive function, memory loss | 1 | 2 | 2 | 3[a] | 3 | 3 | 2 | 6[b] | 15 | 9 |
| Concentration improved | | | 1 | | | | | | 1 | 0 |

At interview post-DBS experience ranged from [a] 14-129, [b] 13-61 months

Other comments valid at interview that have been grouped under this heading include vagueness, confusion, loss of memory (especially with words), loss of verbal skills, loss of numeracy skills, thinking patterns "scrambled", poor financial decisions and dementia.

In contrast, one recipient reported improved concentration at about 12 months following the procedure.

<u>Sleep</u>

There were a number of references to changes in sleep quality (Table 11). Over the first 12 months following the procedure there was increasing mention of poorer sleeping experiences. Poorer sleep was mentioned at some time in the post-DBS period in 12 of 52 cases (23%) and at the time of interview in 6 of 52 cases (12%).

| Table 11- Number of Comments About Sleep | | | | | | | | | | |
|---|---|---|---|---|---|---|---|---|---|---|
| | DBS recipient | | | | Carer | | | | Cases total | Cases at Int |
| **Months after initial DBS procedure** | 1 | 6 | 12 | Int | 1 | 6 | 12 | Int | | |
| **Comments** | | | | | | | | | | |
| Sleeping worse | 1 | 1 | 4 | 4$^a$ | 1 | 2 | 4 | 4$^b$ | 12 | 6 |
| Sleeping better | 4 | 3 | | 2$^c$ | 5 | | | | 8 | 2 |
| Sleeping more | | | | | | | | 1$^d$ | 1 | 1 |
| Having nightmares | | 1 | 1 | 1$^e$ | | | 1 | 1$^e$ | 1 | 1 |
| Having vivid dreams | 1 | | 1 | | | | | | 2 | 0 |
| Better at turning in bed | 1 | | | | 1 | | | | 2 | 0 |

At interview post-DBS experience ranged from $^a$ 11-129, $^b$ 14-219, $^c$ 46-60, $^d$ 14, $^e$ 12 months

Over the first 12 months following the procedure there was increasing mention of poorer sleeping experiences. Poorer sleep was mentioned at some time in the post-DBS period in 12 of 52 cases (23%) and at the time of interview in 6 of 52 cases (12%).

In contrast there were a number of reports of better sleeping for the first month or so following the procedure. Similar reports of this nature came from the recipient and observed by the carer. Reports of this nature related to at least some period following DBS in 8 of 52 cases (15%) and valid at the time of interview in 2 of 52 cases (4%).

Dreams or nightmares were mentioned in a few cases. Being able to turn more easily in bed was also mentioned in two cases.

Mood and Behaviour

There were many features mentioned that have been grouped as mood and behaviour (Table 12). The most common report in this grouping, largely from carers, was for a more relaxed and happier DBS recipient. There were 9 in 52 cases (17%) in which this was mentioned at some time in the post-DBS period, and 4 in 52 cases (8%) when this was mentioned valid at the time of interview.

Tiredness or fatigue was also reported in a number of cases – in 7 of 52 cases (13%) at least at some time post DBS and in 4 of 52 cases (8%) valid at the time of interview.

| Table 12– Number of Comments About Mood and Behaviour | | | | | | | | | | |
|---|---|---|---|---|---|---|---|---|---|---|
| | DBS recipient | | | | Carer | | | | Cases total | Cases at Int |
| **Months after initial DBS procedure** | 1 | 6 | 12 | Int | 1 | 6 | 12 | Int | | |
| **Comments** | | | | | | | | | | |
| More relaxed, happier, brighter | 2 | | | 1$^a$ | 5 | 3 | 2 | 3$^b$ | 9 | 4 |
| Tired, fatigued | 1 | 3 | 3 | 2$^c$ | | 1 | 1 | 2$^d$ | 7 | 4 |
| Obsessive, compulsive, impulsive, uninhibited behaviour | | | | | 3 | 3 | | 2$^c$ | 6 | 2 |
| Apathy, lethargy worse | 1 | 1 | 1 | | 1 | 3 | 2 | 2$^e$ | 4 | 2 |
| Stays around home | | | | | | | | 2$^f$ | 2 | 2 |
| More confident | 2 | 1 | 1 | 1$^g$ | 1 | | 2 | | 4 | 1 |
| More caring nature | | | | | 1 | | | 1$^h$ | 2 | 1 |
| Emotional | | | 1 | 1$^i$ | | | | | 1 | 1 |
| Less depressed | | | | | 1 | 1 | 1 | 1$^j$ | 1 | 1 |
| More anxious | | | | | | | | 1$^k$ | 1 | 1 |
| Hallucinations | | | 1 | 1$^l$ | | | | | 1 | 1 |

| | | | | | | | | | |
|---|---|---|---|---|---|---|---|---|---|
| Euphoria, elation | 5 | 1 | | | 3 | 1 | | 7 | 0 |
| Mood swings | | 1 | | | 2 | 1 | 1 | 3 | 0 |
| Aggressive, intolerant | | | 1 | | 1 | | 1 | 2 | 0 |
| Less caring nature | | | | | 1 | 1 | | 2 | 0 |
| Less anxious | | | | | 2 | | | 2 | 0 |
| Hilarity | | 1 | | | | | | 1 | 0 |
| Depressed | | 1 | | | | 1 | 1 | 1 | 0 |
| Apathy, lethargy better | | 1 | | | | | | 1 | 0 |
| Introverted | | 1 | | | | | | 1 | 0 |
| More stable mood | | | 1 | | | | | 1 | 0 |
| More independent | | | | | 1 | | | 1 | 0 |
| Hyperactive | 1 | 1 | | | | | | 1 | 0 |

At interview post-DBS experience ranged from [a] 60, [b] 12-23, [c] 12-13, [d] 15-61, [e] 37-60, [f] 12-84, [g] 97, [h] 22, [i] 27, [j] 12, [k] 129, [l] 27 months

Behaviour that could be described as obsessive, compulsive, impulsive or uncharacteristically uninhibited was reported by a number of carers, but by no recipients. Behaviour of this nature was reported in 6 or 52 cases (12%) at some time post-DBS and in 2 of 52 cases (4%) at the time of interview.

Apathy or lethargy was mentioned by both recipients and carers in a number of cases – noticed in 4 of 52 (8%) at some time following DBS and in 2 of 52 (4%) valid at the time of interview. Another perhaps related series of comments implied decreasing wish for socialisation (was happy to just stay around the home). Comments of this nature were reported just by carers in 2 of 52 cases (4%) valid at the time of interview.

There were many more comments grouped under the mood and behaviour heading (Table 12) but reported in relatively few cases. One that requires further comment is the cluster "Euphoria, elation". The reports of this nature were noteworthy in two respects – they were observations from both recipients and carers and all were restricted to the first few months following the procedure. In that short period this topic was reported in a total of 7 of 52 cases (13%), but none valid at the time of interview.

Pain

Pain was reported by a number of participants. This topic was often difficult to assess for the purposes of this survey. The important questions were whether the perceived pain was either a feature of Parkinson's disease or DBS, because many participants had other health problems in addition to PD (comorbidities). The responses in Table 13 reflect their assessment of relevancy to this survey.

| Table 13 - Number of Comments About Pain | | | | | | | | | | |
|---|---|---|---|---|---|---|---|---|---|---|
| | DBS recipient | | | | Carer | | | | Cases total | Cases at Int |
| Months after initial DBS procedure | 1 | 6 | 12 | Int | 1 | 6 | 12 | Int | | |
| Comments | | | | | | | | | | |
| Headaches | 1 | | | 1[a] | | | | | 2 | 1 |
| Pain over scalp | 1 | 1 | | 1[b] | | | | | 2 | 1 |
| Back pain | | | | 1[c] | | | | | 1 | 1 |
| Leg or hip pain | | 2 | | | | 1 | | | 3 | 0 |
| Chest pain | 1 | | 1 | | | | | | 2 | 0 |
| Neck pain | 1 | 2 | 2 | | | 1 | | | 2 | 0 |

At interview post-DBS experience ranged from [a] 17, [b] 11, [c] 27 months

Most reports of pain were understandably from the recipient, but in two cases carers reported observing the effects. Various sites were mentioned for the pain but headaches, pain over the scalp and back pain were valid at the time of interview, each for 1 in 52 cases (2%). Pain at other sites was reported in a small number of cases, but not active at the time of interview.

Other Non-Motor Symptoms

Comments relating to other non-motor symptoms of Parkinson's disease were relatively few in number (Table 14). Improved eating or weight gain was mentioned in 4 of 52 cases (8%) at some time in the post DBS period, but such comments became fewer with time following DBS with only 1 case (2%) mentioned at the time of interview. In some cases, where weight maintenance had been a problem before DBS, this change was welcomed. Worse incontinence was mentioned in 2 of 52 cases (4%) at some time following the procedure, but only in one case observed by the carer at the time of interview, 61 months following the procedure. Improved ability to smell and taste was reported for the months following the procedure in 2 of 52 cases (4%), but none at the time of interview.

| Table 14– Other Comments Related to Non-motor Symptoms | | | | | | | | | | |
|---|---|---|---|---|---|---|---|---|---|---|
| | DBS recipient | | | | Carer | | | | Cases total | Cases at Int |
| **Months after initial DBS procedure** | 1 | 6 | 12 | Int | 1 | 6 | 12 | Int | | |
| **Comments** | | | | | | | | | | |
| Weight gain, eating better | | 2 | 2 | | 1 | 1 | 1 | 1[a] | 4 | 1 |
| Incontinence worse | | 1 | | | | | | 1[b] | 2 | 1 |
| Smell and taste improved | 1 | 1 | 1 | | | 1 | | | 2 | 0 |
| Nauseous | 1 | | | | | | | | 1 | 0 |
| Feeling unwell | | | 1 | | | | | | 1 | 0 |
| Sight improved | 1 | 1 | 1 | | | | | | 1 | 0 |

At interview post-DBS experience ranged from [a] 129, [b] 61 months

## 4. PD Management After DBS

There were a number of reports that were related to PD management following DBS (Table 15).

Both recipients and carers reported a desire for more information to be provided to them about DBS prior to the procedure. Reports of this nature were received in 7 of 52 cases (13%).

Another group, predominantly carers, reported a desire for more information and support following DBS. This applied in 5 of 52 cases (10%).

At the time of interview there were reports of the benefits of DBS wearing off in 5 of 52 cases (10%). Those recipients reporting this had DBS at some time in the range 24-129 months prior to interview.

| Table 15- Number of Comments About PD Management After DBS | | | | | | | | | | |
|---|---|---|---|---|---|---|---|---|---|---|
| | DBS recipient | | | | Carer | | | | Cases total | Cases at Int |
| **Months after initial DBS procedure** | 1 | 6 | 12 | Int | 1 | 6 | 12 | Int | | |
| **Comments** | | | | | | | | | | |
| Would like more information prior to DBS | | | | 4[a] | | | | 5[b] | 7 | 7 |
| Would like more information and support following DBS | | | | 2[c] | | | | 5[d] | 5 | 5 |
| Benefits of DBS wearing off | | | | 4[e] | | 1 | | 2[f] | 5 | 5 |

| Symptom | | | | | | | | | |
|---|---|---|---|---|---|---|---|---|---|
| Involuntary movements, dyskinesia worse | | 1 | 1 | 1[g] | | 1 | 1 | 1[h] | 2 | 2 |
| Difficulty in finding correct DBS settings/ frequent adjustments | | | | | 1 | | 1 | 1[i] | 2 | 1 |
| No stable management of PD symptoms | | 1 | 2 | 1[i] | 1 | | | | 2 | 1 |
| High levels of PD medication needed | | | | 1[j] | | | | | 1 | 1 |
| Totally dependent | | | | | | | | 1[k] | 1 | 1 |
| Sensations in brain or body caused by DBS or settings change | 9 | 5 | 1 | | | | | | 9 | 0 |
| Better "on" time | 2 | | | | | | | | 2 | 0 |
| Decrease in medication | 1 | 1 | | | | | | | 1 | 0 |

At interview post-DBS experience ranged from [a] 13-22, [b] 13-85, [c] 22-74, [d] 15-74, [e] 24-129, [f] 60-66, [g] 14, [h] 12, [i] 13, [j] 73, [k] 74 months

Other comments relating to the management of PD post-DBS were wide ranging but any particular topic was only mentioned in a relatively small number of cases at interview (Table 15). One that had more numerous mentions early in the post-DBS period related to sensations experienced caused by DBS or adjustment of DBS settings. A number of recipients reported sensations when the settings were changed. Examples included sensations like a current surge or jolt in the brain, tingling in the brain or elsewhere in the body or transient "fogginess" in the brain, but these sensations were transient - all such comments were restricted to the first few months following the procedure. None was mentioned at the time of interview.

## Part B – Perceived Changes in Symptoms Compared With Conditions Prior To DBS

In this section participants were asked to rate experiences of 32 symptoms according to the scale shown in Table 16.

Participants were asked to provide numeric scores according to the table shown above for the times following the DBS procedure of 1 month, 6 months, 12 months and at the time of interview. In those few cases where participants had had repeat DBS procedures, timing was measured from the first implantation. Where interviews were conducted close to the first anniversary of the procedure, the same responses were recorded under the headings "after 12 months" and "at interview".

Not all participants provided scores for every question. In the following figures, $N_R$ = *number of recipients that provided a score of 0-5,* and $N_C$ = *number of carers who provided a score of 0-5.*
Results are shown in two columns. The left column shows the scores provided by the combined pool of DBS recipients and carers and are therefore based on the total number $(N_R + N_C)$ of responses. The right column shows the difference in scores between DBS recipients and carers intended to identify symptoms that were assessed differently by DBS recipients and carers.

| Table 16– Scoring Schedule | |
|---|---|
| **Score** | **Meaning** |
| 5 | Much better after DBS |
| 4 | Better after DBS |
| 3 | About the same as before DBS |
| 2 | Worse after DBS |
| 1 | Significantly worse after DBS |
| 0 | Not applicable (e.g. did not exhibit this symptom up to this point in time) |
| Blank | Unwilling or unable to answer |

Care is needed in interpreting data relating to differences in response – for a number of cases different numbers of recipients and carers responded, and there were a number of unmatched recipients and carers. Nevertheless the numbers presented do give some indication of any differences in assessment by recipients as a group and by carers as a group. *Number (R-C)* is, for each score, the number of recipients minus the number of carers giving this score. If positive, this indicates more recipients than carers gave that response. If negative, more carers than recipients gave that response.

1. **Motor Symptoms**

People living with PD can experience a wide range of symptoms, but not all known symptoms. Scores of "0" indicated that the symptom in question was not relevant (applicable) to the case reported on by that participant. For motor symptoms considered here these "not applicable" responses are shown in the table below. The figures shown apply to the initial assessment of participants, i.e. in the period immediately following DBS (1 month). In some cases the symptom **did** become "applicable" at some time after DBS, in which circumstances the "not applicable" responses decreased over time as can be seen in some of the following Figures. Comments following in this section refer to those who did experience the symptoms i.e. did return responses in the range 1-5.

Results for the different symptoms are presented in bar chart form in the order of decreasing applicability as indicated in the Table 17. The most common symptom on this list is *handwriting*. All participants who responded to this question (90 in total) indicated this was applicable to their situation (i.e. none described it as "not applicable"). In contrast *freezing* wasn't relevant for 45 of the 88 participants who responded to this symptom.

| Table 17– Relevance (Applicability) of Motor Symptoms to Individual Cases | | | |
|---|---|---|---|
| Symptom | Results | Reported initially as "Not Applicable" | |
| | Figure number | Participant count | Percentage |
| Handwriting | 2 | 0/90 | 0 |
| Difficulty walking | 3 | 4/88 | 5 |
| Speech | 4 | 7/89 | 8 |
| Postural stability | 5 | 11/90 | 12 |
| Slowness or lack of movement | 6 | 13/88 | 15 |
| Facial expression | 7 | 15/79 | 19 |
| Difficulty standing from a chair | 8 | 19/90 | 21 |
| Difficulty turning in bed | 9 | 15/60 | 25 |
| Tremor | 10 | 24/91 | 26 |
| Stiffness and rigidity | 11 | 24/88 | 27 |
| Swallowing | 12 | 38/88 | 43 |
| Freezing | 13 | 45/88 | 51 |

Results in Table 17 show that there is wide variation in the relevance (applicability) of many motor symptoms. The subsequent discussion of results in this section for these motor symptoms is restricted to those responses that indicated that the symptom in question was applicable i.e. for which a score was given in the range 1-5.

All scores in the range 1-5 were averaged at each of the post-DBS time intervals chosen for this study. The results are listed in Table 18 in the same order as Table 17 (i.e. based on the relevance of symptoms in order most relevant to least relevant). These averages give an indication of the average outcomes perceived by the participants. A more graphic indication of the differences in scores, and variation over time is provided in the following Figures (2-13), also shown in the same order as Table 17.

| Table 18– Average Scores (1 – 5) for Motor Symptoms Listed in Decreasing Order of Applicability | | | | | | | | |
|---|---|---|---|---|---|---|---|---|
| Symptom | DBS + 1 month | | DBS + 6 months | | DBS + 12 months | | At interview | |
|  | R | C | R | C | R | C | R | C |
| Handwriting | 3.29 | 3.13 | 3.04 | 2.96 | 2.84 | 2.86 | 2.56 | 2.60 |
| Difficulty walking | 3.70 | 3.77 | 3.58 | 3.52 | 3.26 | 3.29 | 2.95 | 2.86 |
| Speech | 3.28 | 3.10 | 3.12 | 3.00 | 2.72 | 2.49 | 2.49 | 2.34 |
| Postural stability | 3.38 | 3.54 | 2.90 | 2.95 | 2.68 | 2.92 | 2.65 | 2.71 |
| Slowness or lack of movement | 3.81 | 3.92 | 3.72 | 3.62 | 3.68 | 3.62 | 3.35 | 3.15 |
| Facial expression | 3.80 | 3.94 | 3.68 | 3.71 | 3.55 | 3.64 | 3.35 | 3.43 |
| Difficulty standing from a chair | 3.54 | 3.67 | 3.34 | 3.41 | 3.21 | 3.23 | 2.86 | 2.87 |
| Difficulty turning in bed | 3.78 | 3.86 | 3.65 | 3.67 | 3.48 | 3.62 | 3.43 | 3.50 |
| Tremor | 4.49 | 4.31 | 4.43 | 4.28 | 4.42 | 4.20 | 4.26 | 4.18 |
| Stiffness and rigidity | 3.87 | 3.91 | 3.71 | 3.67 | 3.66 | 3.52 | 3.31 | 3.18 |
| Swallowing | 3.12 | 2.75 | 3.08 | 2.76 | 3.00 | 2.70 | 2.79 | 2.40 |
| Freezing | 3.85 | 3.91 | 3.37 | 3.59 | 3.00 | 3.32 | 2.72 | 2.75 |

Participants reported a general improvement to many motor symptoms following DBS, but there were significant differences depending on the symptom.

<u>Significant and sustained improvements</u>

Of all symptoms surveyed *tremor* (Table 18 and Fig. 10) showed the strongest positive improvement following DBS. Most participants indicated that this was much better after DBS and that the improvements were sustained over time with average scores of 4.42/4.20 (R/C) 12 months after DBS and 4.26/4.18 (R/C) at the time of interview.

<u>Sustained improvements</u>

Other symptoms that on average showed improvement included *slowness or lack of movement* (Fig. 6), *facial expression ("Parkinson's face")* (Fig. 7), *difficulty turning in bed* (Fig. 9) and *stiffness and rigidity* (Fig. 11). Symptoms in this group were reported by many participants to be better or much better following DBS with average scores above 3 through to the time of interview.

<u>Improvements but not sustained</u>

Another group of symptoms, including *difficulty walking* (Fig. 3), *difficulty standing from a chair* (Fig. 8) and *freezing* (Fig. 13), showed initial improvement, but more decline in improvement with time such that at the time of interview the average scores were less than 3.

Scores for *speech* and *postural stability* showed that on average these symptoms showed some improvement in the few months following the procedure but that this improvement wasn't sustained.

In the case of *speech* (Fig. 4) the largest number of responses in the first 12 months following DBS was for a score of 3 (much the same after DBS) but scores of either 1 or 2 (worse after DBS) increased with time following the procedure such that at 12 months the average scores were 2.72/2.49 (R/C) and at interview, the largest number of responses was for a score of 2 with overall averages of 2.49/2.34 (R/C). These results were in broad agreement with those in Part A in which there was frequent mention of poorer speech (Table 5).

As discussed earlier in relation to Part A of the survey, balance problems and falls were common experiences. In Part B postural stability was assessed by participants in terms of propensity for falling. Improvements in *postural stability* (Fig. 5) were reported by some in the first month or so following DBS, but often this was not sustained. By the end of the first year following DBS, the score with the largest number of responses was 2 and the average scores 2.68/2.92 (R/C) indicating that at this time this symptom was worse than before DBS for many. The number reporting a score of 2 increased monotonically with time following DBS such that at the time of interview the average scores had reduced to 2.65/2.71 (R/C).

Little or no improvements

Some symptoms appear to show little or no improvement with DBS. These include *handwriting* (Fig. 2) and *swallowing* (Fig. 12). Most participants for whom these symptoms were relevant gave a score of 3 (much the same after DBS) for these symptoms with some evidence of worsening by the time of the interview. The average responses 12 months after DBS were 2.84/2.86 (R/C) for handwriting and 3.00/2.70 (R/C) for swallowing, and even lower at the time of interview.

Comparison of responses, recipients and carers

Allowing for the fact that, for many symptoms, there were different numbers of recipients and carers providing scores, recipients and carers were generally consistent in the scores given. However there were some notable differences with some symptoms. With *tremor* (Fig. 10) there were some differences reported in the level of improvement – more carers than recipients gave a score of 4 but more recipients than carers gave a score of 5. Early in the post-DBS period more carers than recipients reported scores of 5 (much better after DBS) for the symptoms of *difficulty walking* (Fig. 3), *slowness or lack of movement* (Fig. 6), *difficulty standing from a chair* (Fig. 8) and *freezing* (Fig. 13) but this difference was not sustained at later times. More carers than recipients gave scores of either 4 or 5 for *facial expression* (Fig. 7).

**Motor Symptom Responses in Order of Decreasing Applicability**

**(a) Combined scores –
 DBS recipients plus carers**

**(b) Differences in scores –
 (DBS recipients – carers)**

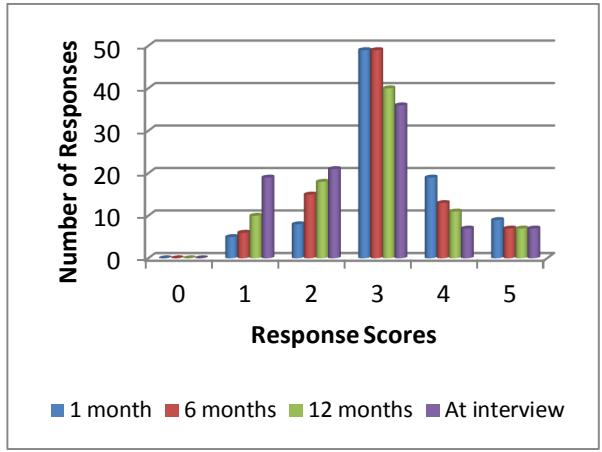
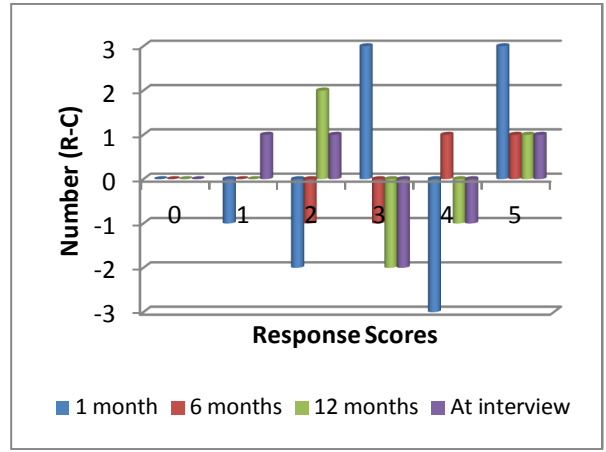

**Fig. 2 Handwriting** $(N_R = 45, N_C = 45)$

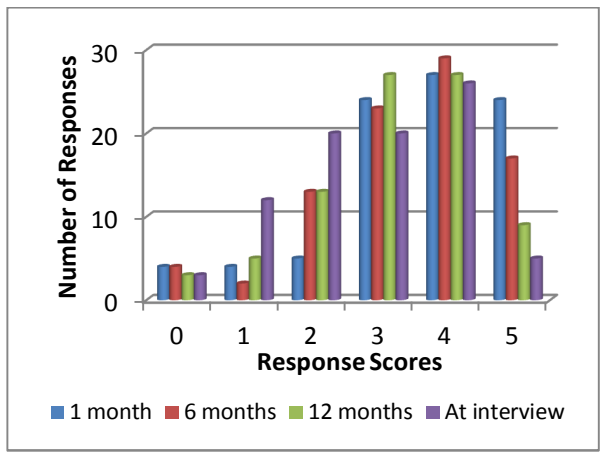
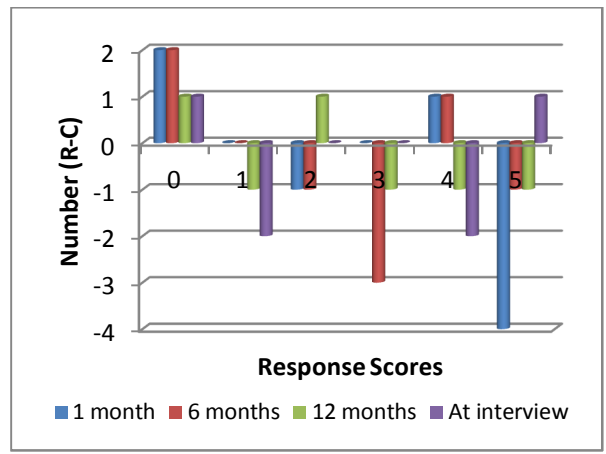

**Fig. 3 Difficulty Walking** $(N_R = 43, N_C = 45)$

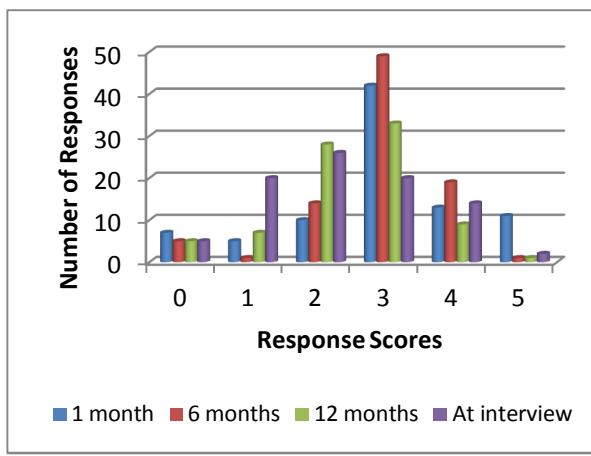
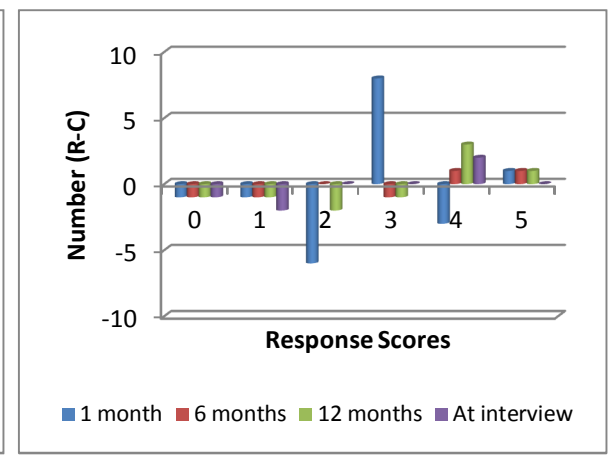

**Fig. 4 Speech** $(N_R = 44, N_C = 45)$

**Motor Symptom Responses in Order of Decreasing Applicability (cont.)**

**(a) Combined scores –**
   **DBS recipients plus carers**

**(b) Differences in scores –**
   **(DBS recipients – carers)**

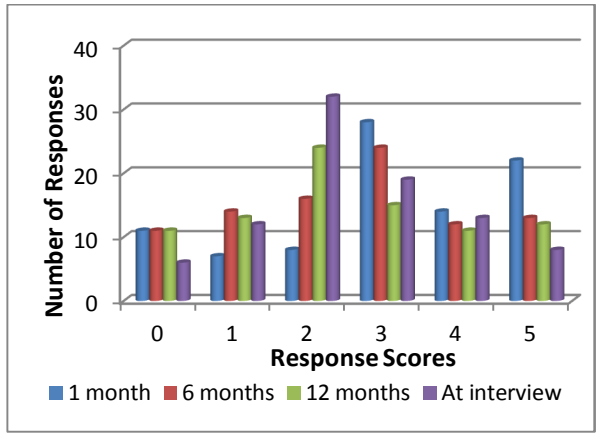
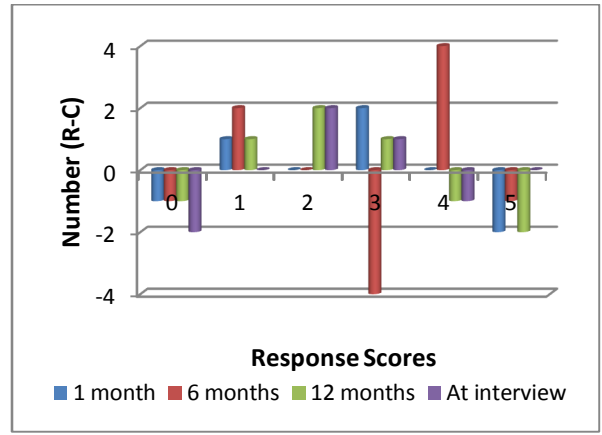

**Fig. 5 Postural Stability** *($N_R$ = 45, $N_C$ = 45)*

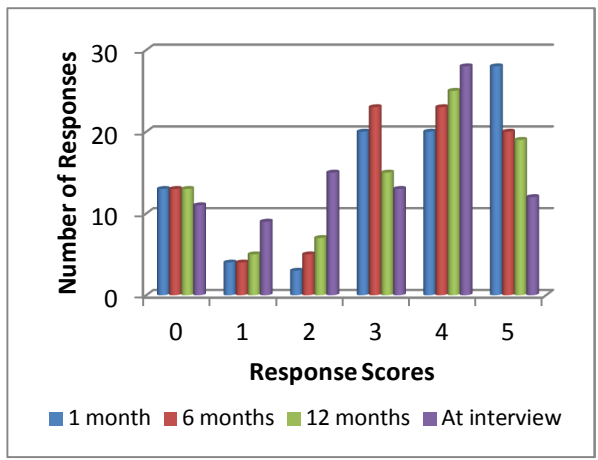
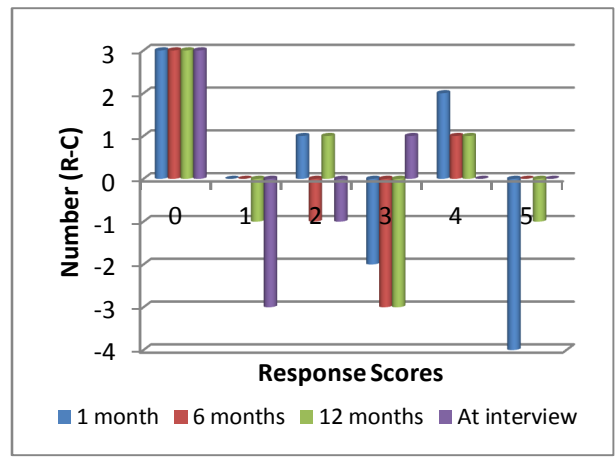

**Fig. 6 Slowness or Lack of Movement** *($N_R$ = 44, $N_C$ = 44)*

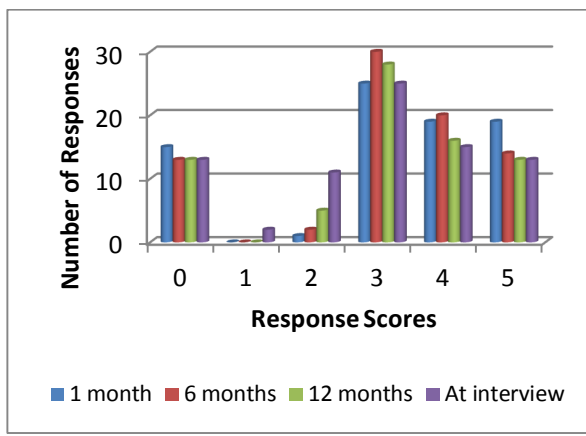
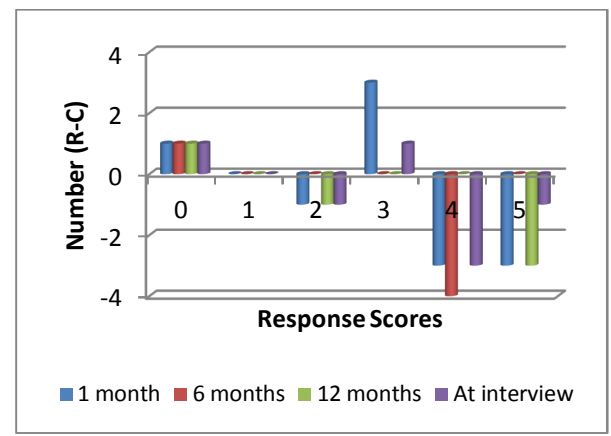

**Fig. 7 Facial Expression** *($N_R$ = 38, $N_C$ = 41)*

**Motor Symptom Responses in Order of Decreasing Applicability (cont.)**

**(a) Combined scores –
DBS recipients plus carers**

**(b) Differences in scores –
(DBS recipients – carers)**

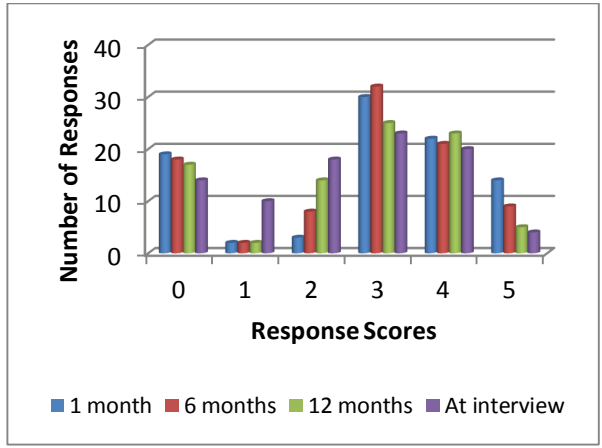
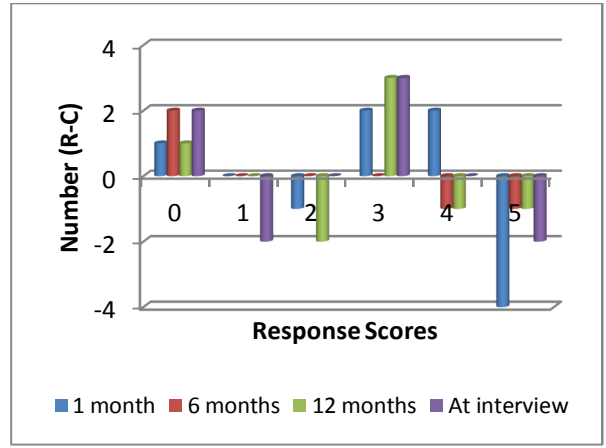

**Fig. 8 Difficulty Standing From Chair** *($N_R$ = 45, $N_C$ = 45)*

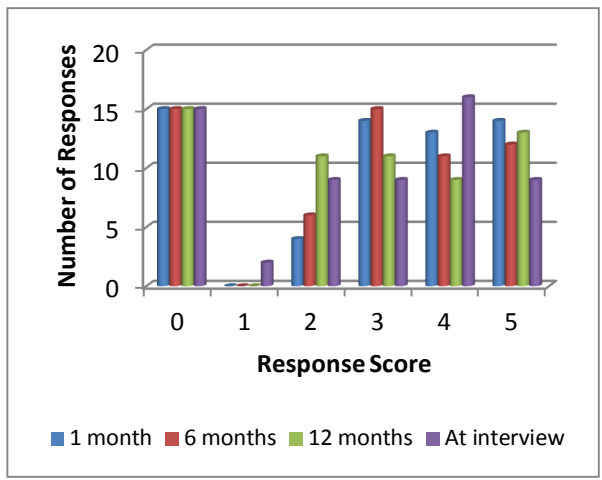
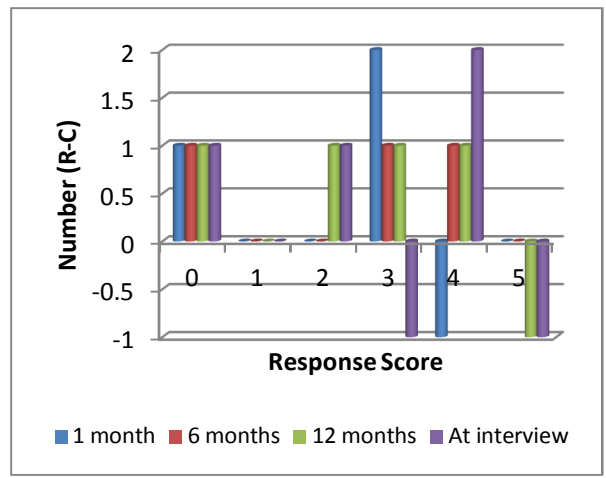

**Fig. 9 Difficulty Turning in Bed** *($N_R$ = 31, $N_C$ = 29)*

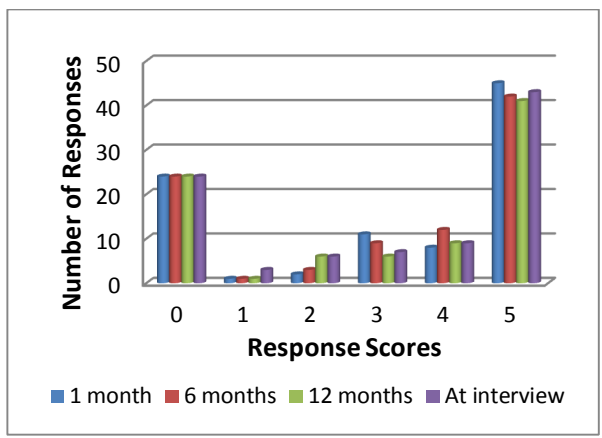
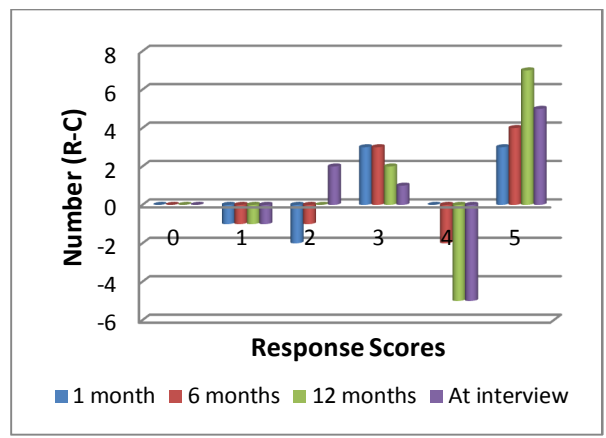

**Fig. 10 Tremor** *($N_R$ = 47, $N_C$ = 44)*

## Motor Symptom Responses in Order of Decreasing Applicability (cont.)

**(a) Combined scores –**
   **DBS recipients plus carers**

**(b) Differences in scores –**
   **(DBS recipients – carers)**

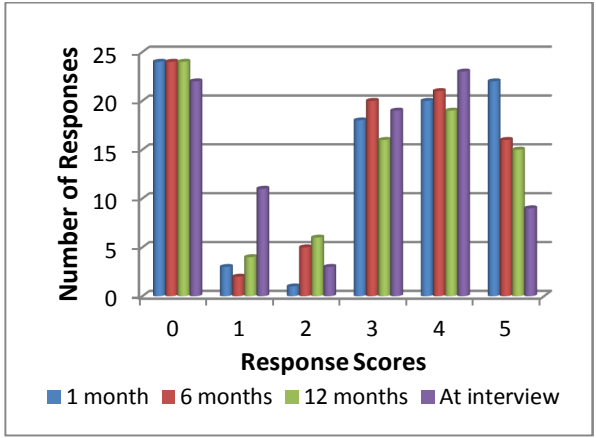
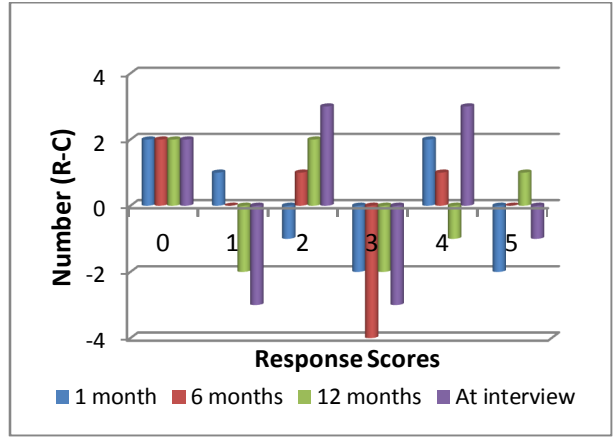

**Fig. 11 Stiffness and Rigidity** *($N_R$ = 44, $N_C$ = 44)*

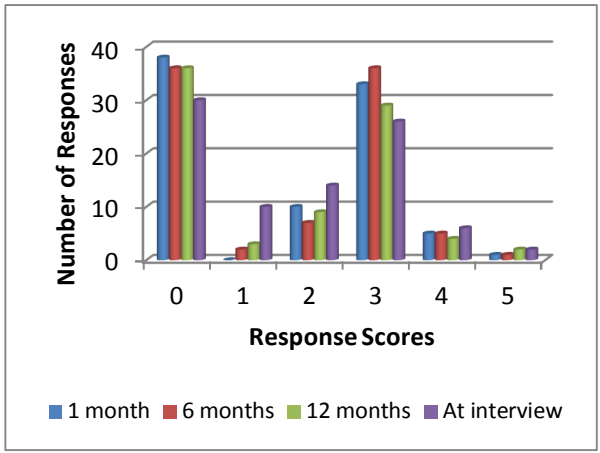
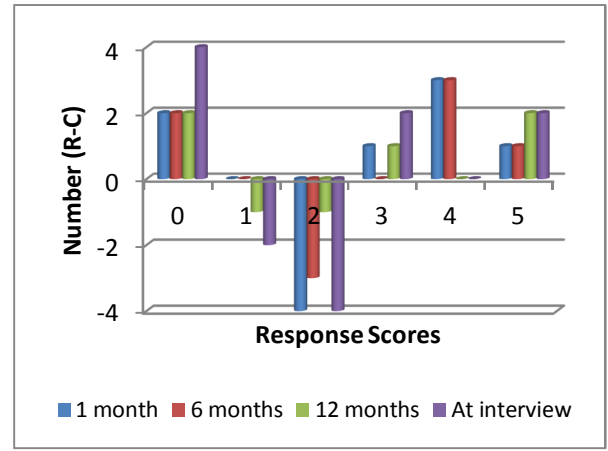

**Fig. 12 Swallowing** *($N_R$ = 45, $N_C$ = 43)*

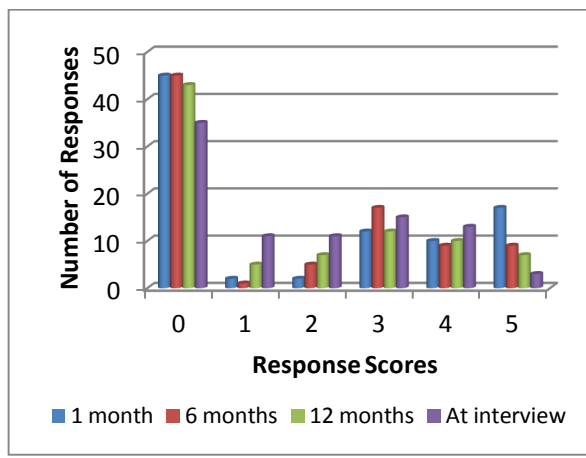
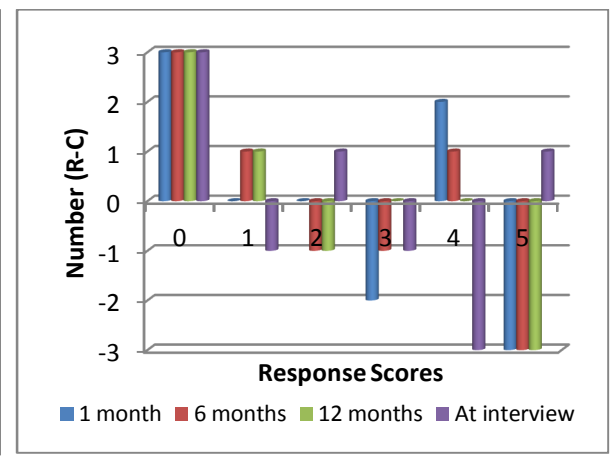

**Fig. 13 Freezing** *($N_R$ = 44, $N_C$ = 44)*

## 2. Non-Motor Symptoms

For non-motor symptoms considered in this section, the numbers of "not applicable" responses are shown in the table below. As for the motor symptoms discussed in the previous section, the figures shown in the table apply to the assessment of participants in the period immediately following DBS (1 month). In many cases the symptom **did** become "applicable" at some time after DBS, in which cases the "not applicable" responses decreased over time as can be seen in some of the following Figures. Comments following in this section refer to those who did experience the symptoms i.e. did return responses in the range 1-5. Results for the different symptoms are presented in bar chart form in the order of decreasing applicability as indicated in Table 19 shown below.

| Table 19– Relevance (Applicability) of Non-Motor Symptoms to Individual Cases | | | |
|---|---|---|---|
| Symptom | Results Figure number | Reported initially as "Not Applicable" | |
| | | Participant count | Percentage |
| Sleep quality | 14 | 3/88 | 3 |
| Fatigue | 15 | 9/89 | 10 |
| Cognitive function, memory and reasoning ability | 16 | 17/88 | 19 |
| Sense of Smell | 17 | 20/83 | 24 |
| Mood and behaviour | 18 | 17/60 | 28 |
| Constipation | 19 | 22/74 | 30 |
| Anxiety | 20 | 31/85 | 36 |
| Severity of on/off periods | 21 | 35/82 | 43 |
| Dystonia | 22 | 43/84 | 51 |
| Apathy | 23 | 46/89 | 52 |
| Vision | 24 | 44/83 | 53 |
| Depression | 25 | 47/87 | 54 |
| Dyskinesias | 26 | 45/84 | 54 |
| Pain | 27 | 42/77 | 55 |
| Incontinence | 28 | 46/81 | 57 |
| Restless leg | 29 | 49/86 | 57 |
| Thermoregulatory dysfunction | 30 | 49/60 | 82 |
| Hallucinations or delusions | 31 | 72/85 | 85 |

It can be seen from Table 19 that there is wide variation in the relevance (applicability) of many symptoms. Dystonia, apathy, vision problems, depression, dyskinesias, pain, incontinence, restless leg, thermoregulatory dysfunction, hallucinations or delusions were reported as "not applicable" by more than 50% of participants. Thermoregulatory dysfunction (excessive sweating) and hallucinations or delusions were reported as "not applicable" in over 80% of cases. The subsequent discussion of results in this section for these non-motor systems is restricted to those responses that indicated that the symptom in question was applicable i.e. for which a score was given in the range 1-5.

As for the motor symptoms, all scores in the range 1-5 for the non-motor symptoms were averaged at each of the time intervals chosen for this study. The results are listed in Table 20 and give an indication of the average outcomes perceived by both recipients and carers. Symptoms are listed in the same order as for Table 19. A more graphic indication of the variation of scores over time is provided in the following Figures (14-31), also shown in the order indicated in Table 19.

| Table 20– Average Scores (1 – 5) for Non-Motor Symptoms Listed in Decreasing Order of Applicability | | | | | | | | |
|---|---|---|---|---|---|---|---|---|
| Symptom | DBS + 1 month | | DBS + 6 months | | DBS + 12 months | | At interview | |
| | R | C | R | C | R | C | R | C |
| Sleep quality | 3.56 | 3.43 | 3.58 | 3.57 | 3.59 | 3.53 | 3.47 | 3.55 |
| Fatigue | 3.44 | 3.37 | 3.28 | 3.20 | 3.21 | 3.05 | 2.88 | 2.83 |
| Cognitive function, memory and reasoning ability | 3.14 | 2.80 | 3.03 | 2.80 | 2.97 | 2.73 | 2.72 | 2.49 |
| Sense of Smell | 3.14 | 2.96 | 3.16 | 3.04 | 3.11 | 3.08 | 3.05 | 3.12 |
| Mood and behaviour | 3.71 | 3.05 | 3.52 | 3.05 | 3.14 | 3.05 | 3.43 | 3.32 |
| Constipation | 3.19 | 3.10 | 3.26 | 3.19 | 3.07 | 3.16 | 2.94 | 3.05 |
| Anxiety | 3.59 | 3.52 | 3.48 | 3.33 | 3.43 | 3.07 | 3.10 | 3.00 |
| Severity of on/off periods | 4.00 | 4.08 | 3.91 | 3.87 | 3.90 | 3.57 | 3.73 | 3.52 |
| Dystonia | 4.20 | 4.14 | 4.05 | 4.14 | 3.75 | 3.62 | 3.15 | 3.14 |
| Apathy | 3.10 | 3.23 | 3.05 | 3.18 | 3.15 | 3.10 | 3.14 | 3.04 |
| Vision | 3.25 | 2.95 | 3.20 | 2.95 | 3.28 | 2.82 | 2.95 | 2.50 |
| Depression | 3.71 | 3.42 | 3.43 | 3.05 | 3.23 | 3.00 | 3.23 | 3.00 |
| Dyskinesias | 4.33 | 4.48 | 4.22 | 4.38 | 4.12 | 4.35 | 3.79 | 4.09 |
| Pain | 3.39 | 3.65 | 3.26 | 3.50 | 3.33 | 3.50 | 3.29 | 3.37 |
| Incontinence | 2.89 | 2.88 | 2.83 | 2.88 | 2.76 | 2.81 | 2.55 | 2.50 |
| Restless leg | 3.63 | 3.61 | 3.63 | 3.61 | 3.56 | 3.53 | 3.10 | 3.16 |
| Thermoregulatory dysfunction* | 3.33 | 3.00 | 3.17 | 3.00 | 3.17 | 3.20 | 3.33 | 3.40 |
| Hallucinations or delusions* | 2.75 | 3.20 | 2.88 | 3.40 | 3.38 | 3.40 | 3.30 | 3.80 |

* Notes: The numbers of participants returning scores 1-5 for (a) *thermoregulatory dysfunction* and (b) *hallucinations or delusions* were small – (a) 11 and (b) 13 so the averages shown for these must be treated with added caution.

Significant improvements

The most significant improvements commonly reported by some were for *dyskinesia* (Fig. 26), *dystonia* (Fig. 22) and in improvements in the *severity of on/off periods* (Fig. 21). Improvements in *dyskinesia* were fairly constant over time, with average scores of 4.12/4.35 (R/C) 12 months after DBS and 3.79/4.09 (R/C) at the time of interview. However the improvements with *dystonia* and *severity of on/off* periods showed some decline over time. Both had average scores over 4 at one month following DBS, but by the 12 month mark had dropped to 3.75/3.62 (R/C) in the case of dystonia and 3.90/3.57 (R/C) in the case of severity of on/off periods. At the time of interview there had been further declines in the average scores, more so for dystonia, but average scores were still higher than 3 at that time.

Sustained improvements

Scores for symptoms *sleep quality*, *mood and behaviour* and *pain* showed that, on average, sustained improvements were perceived, at least by recipients. The non-motor symptom of highest applicability was *sleep quality* (Fig. 14). Most reported the same or better sleep quality following DBS, with improvements sustained for many. Average scores were 3.59/3.53 (R/C) 12 months following the procedure and 3.47/3.55 (R/C) at the time of interview. A similar outcome was reported for *mood and behaviour* (Fig. 18) [3.14/3.05 (R/C) at 12 months and 3.42/3.32 (R/C) at interview]. Two features of the results for *mood and behaviour* are especially noteworthy –(a) the generally lower scores given by carers and (b) the improvement in scores from both recipient and carer at the time of interview compared with those at 12 months following DBS. Another symptom that was reported to show sustained improvement was *pain* (Fig. 27) [3.33/3.50 (R/C) at 12 months and 3.29/3.37 (R/C) at interview].

Improvements but not sustained

*Fatigue* (Fig. 15), *anxiety* (Fig. 20), *depression* (Fig. 25) and *restless leg* (Fig. 28) showed improvements after DBS, but these improvements declined such that, at the time of interview, these symptoms were reported to be similar to what they were before DBS (average scores ranged 2.83 – 3.23). In the case of depression it is noticeable that the average scores from carers were consistently lower than those from recipients [3.23/3.00 (R/C) at interview]

Little or no improvements

*Sense of smell* (Fig. 17), *constipation* (Fig. 19), *apathy* (Fig. 23), and *vision* (Fig. 24) did not appear to be much affected by DBS. The most common score was 3 and this did not vary much over time. The results for apathy did not reflect the frequency or worsening of this symptom following DBS reported in the literature (see Fasano [25]). The reasons for this are unclear, but contributing causes might include a lack of understanding of the nature of the symptom, or some stigma attached to it. During the interview process there were a number of responses which were consistent with these as possible causes, but further work would be needed to clarify this.

Few participants responded with scores of 1-5 for *Drenching sweats (thermoregulatory dysfunction)* (Fig. 30) and *hallucinations or delusions* (Fig. 31), but those that did indicated that these symptoms were much the same as before DBS but possibly improving somewhat with time.

About the same or worse

Consistent with the results in Part A of the survey, the group of symptoms *cognitive function, memory and reasoning ability* (Fig. 16) was reported to be about the same or worse after DBS and getting worse over time. One month following DBS the average scores were 3.14/2.80 (R/C) and at interview 2.72/2.49 (R/C). Of note is the average score from carers being always less than 3. Similar results were also reported for *incontinence* (Fig. 28) with average scores ranging 2.89 – 2.50 over the period just after DBS to the time of interview.

Comparison of responses, recipients and carers

There were a number of notable differences between the scoring by DBS recipients and carers for the non-motor symptoms, some of which have already been mentioned above. These differences can be seen in the average scores shown in Table 20 and also (un-averaged) in column (b) of the following Figures. An important qualification to these numbers is that, for many symptoms, there were fewer carers ($N_C$) than recipients ($N_R$) providing scores.

Allowing for this, the data suggest that carers had a more negative assessment for a number of symptoms, at least at some time periods following DBS. These symptoms include *sleep quality* (Fig. 14b), *fatigue* (Fig. 15b), *cognitive function, memory and reasoning ability* (Fig. 16b), *mood and behaviour* (Fig. 18b), *anxiety* (Fig. 20b), *vision* (Fig. 24b) and *depression* (Fig. 25b).

In contrast, carers tended to have more positive assessments for the symptoms of *dyskinesias* (Fig. 26b) and *pain* (Fig. 27).

**Non-Motor Symptom Responses in Order of Decreasing Applicability**

**(a) Combined scores –
DBS recipients plus carers**

**(b) Differences in scores –
(DBS recipients – carers)**

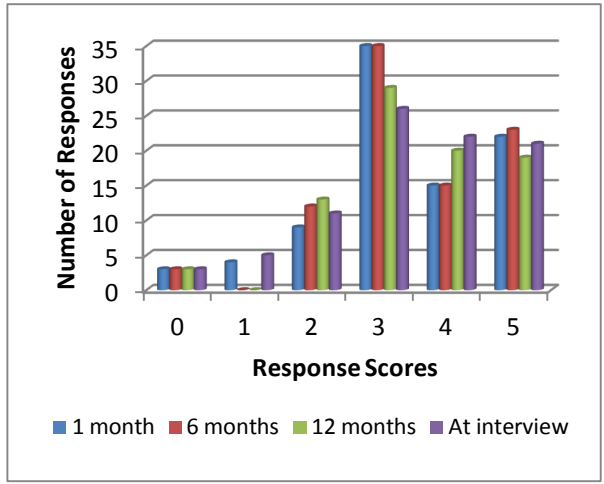
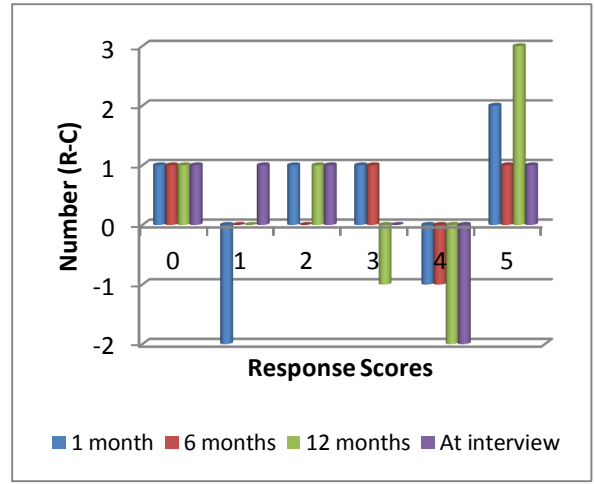

**Fig. 14 Sleep Quality** *($N_R = 45$, $N_C = 43$)*

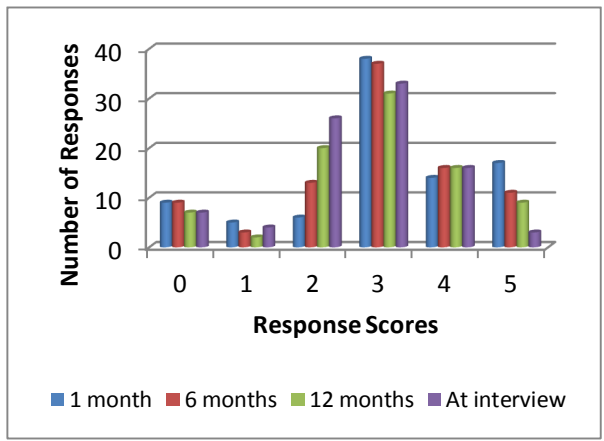
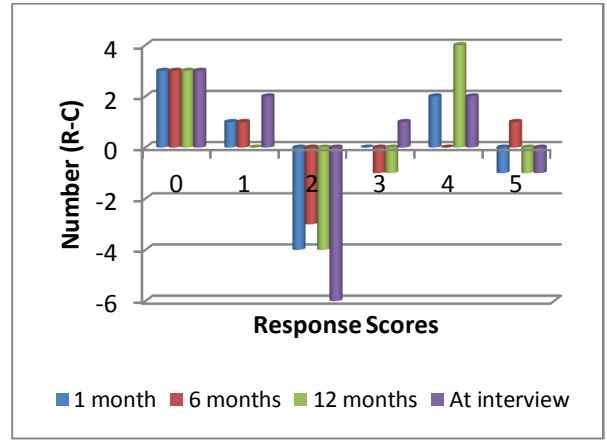

**Fig. 15 Fatigue** *($N_R = 45$, $N_C = 44$)*

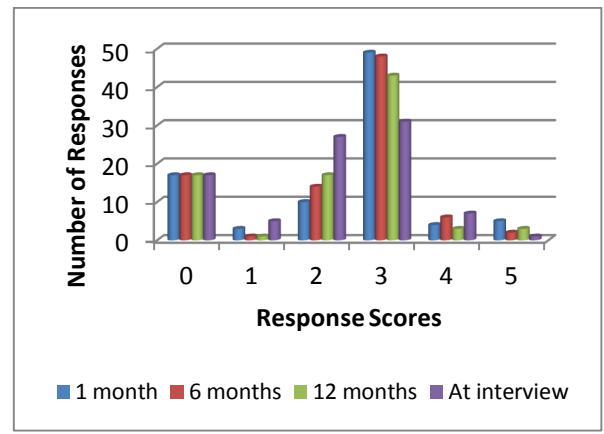
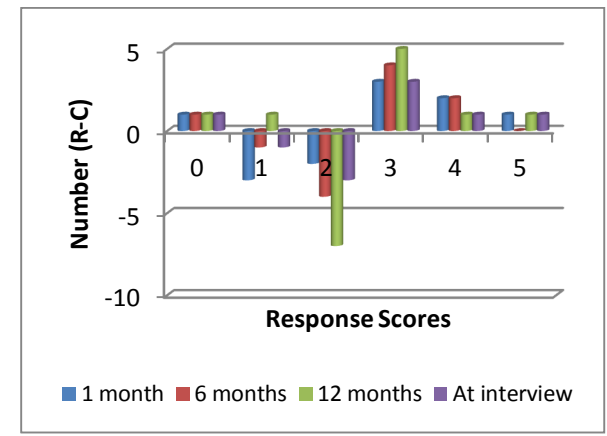

**Fig. 16 Cognitive function, memory and reasoning ability** *($N_R = 45$, $N_C = 43$)*

## Non-Motor Symptom Responses in Order of Decreasing Applicability (cont)

**(a) Combined scores –**
**DBS recipients plus carers**

**(b) Differences in scores –**
**(DBS recipients – carers)**

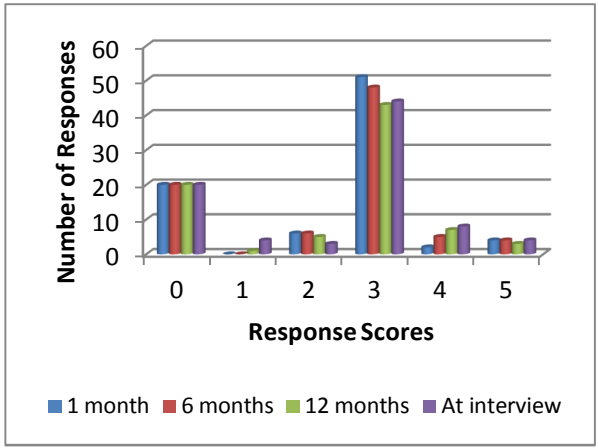
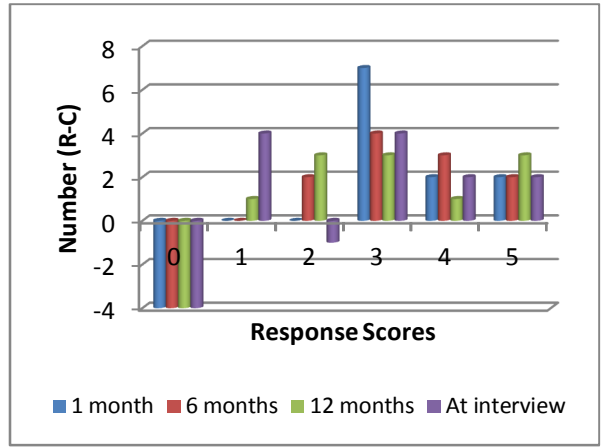

**Fig. 17 Sense of smell** *($N_R$ = 45, $N_C$ = 38)*

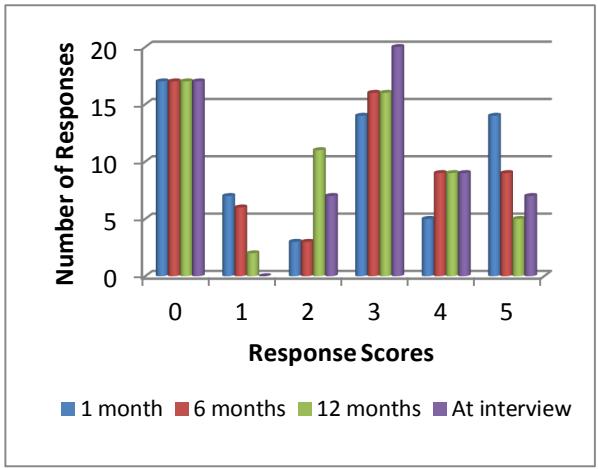
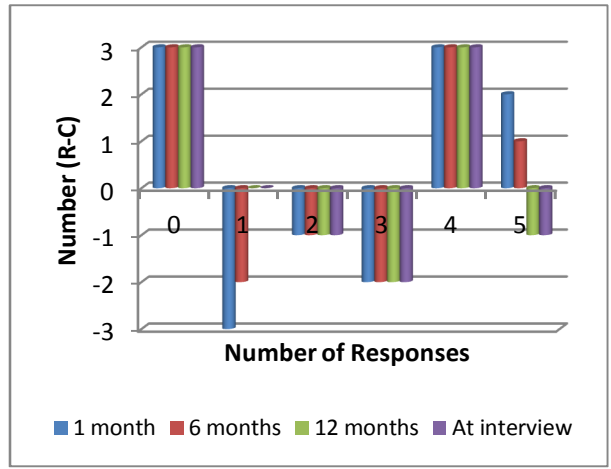

**Fig. 18 Mood and behaviour** *($N_R$ = 31, $N_C$ = 29)*

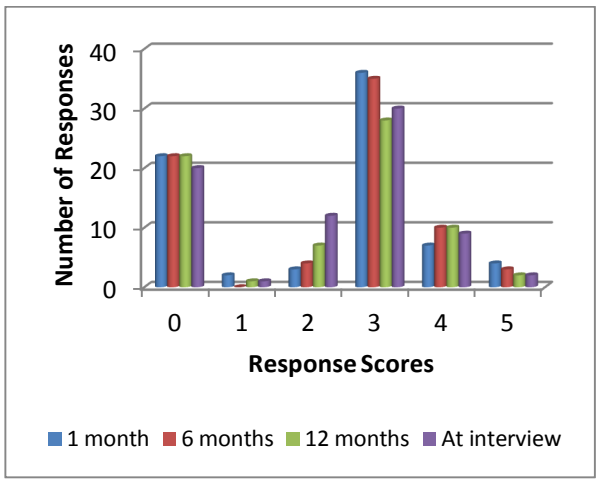
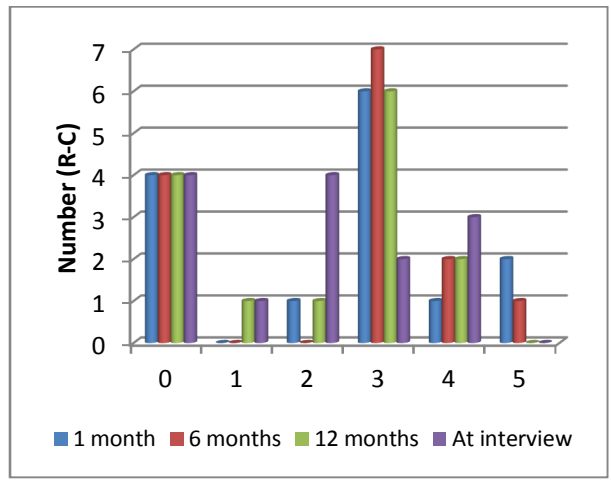

**Fig. 19 Constipation** *($N_R$ = 44, $N_C$ = 30)*

**Non-Motor Symptom Responses in Order of Decreasing Applicability (cont)**

(a) Combined scores –
DBS recipients plus carers

(b) Differences in scores –
(DBS recipients – carers)

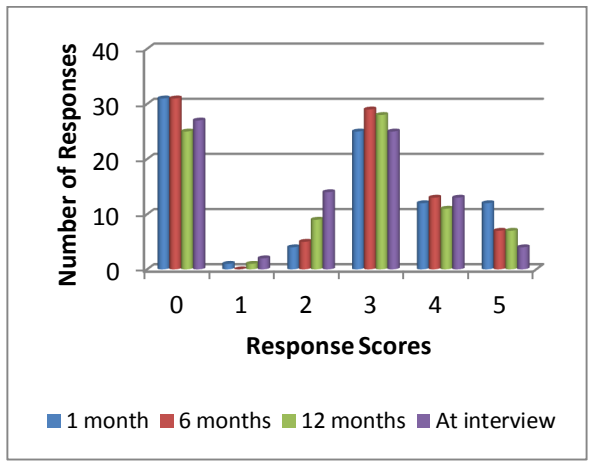
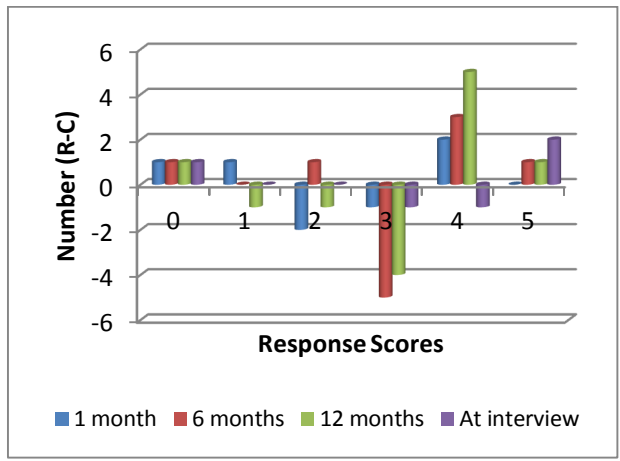

**Fig. 20 Anxiety** *($N_R$ = 43, $N_C$ = 42)*

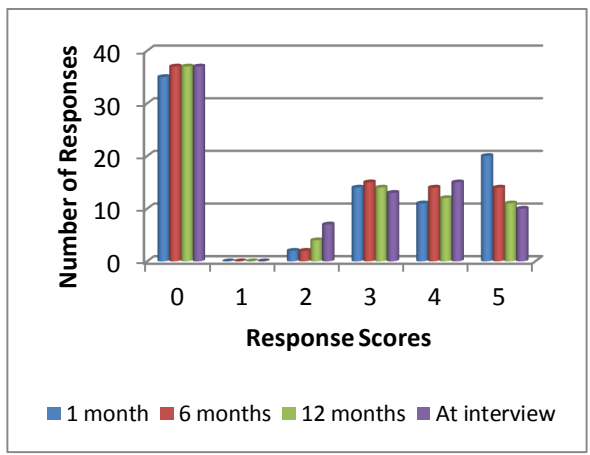
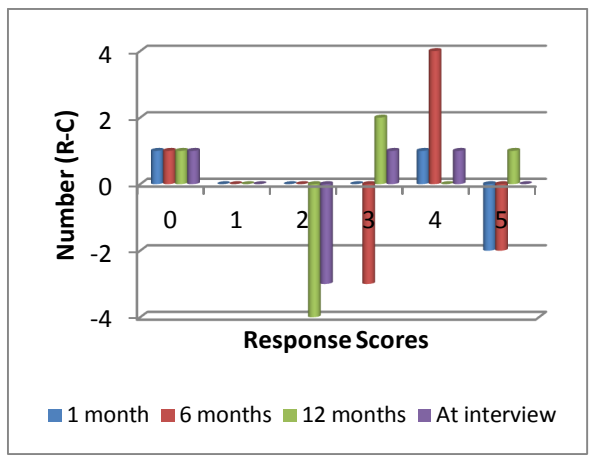

**Fig. 21 Severity of on/off periods** *($N_R$ = 41, $N_C$ = 41)*

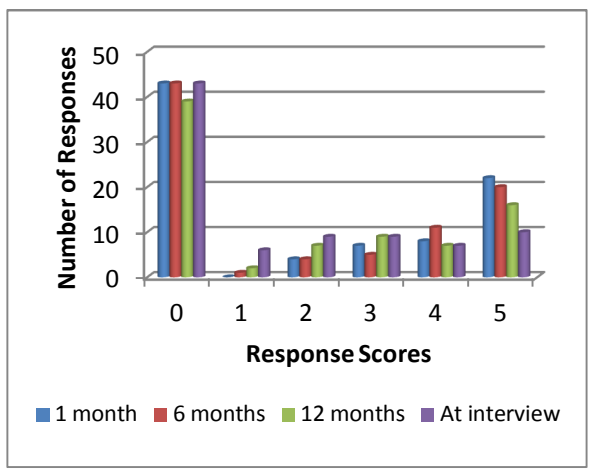
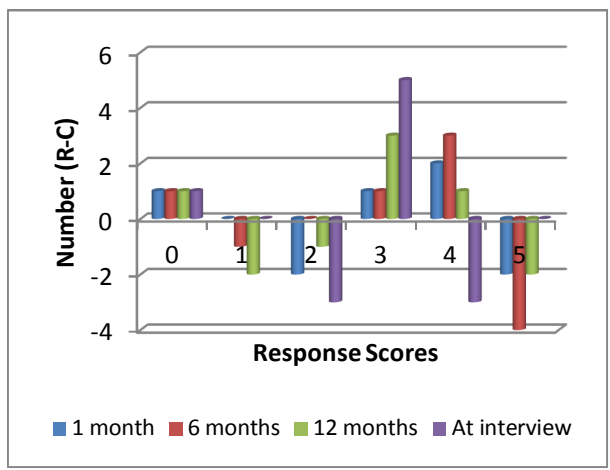

**Fig. 22 Dystonia** *($N_R$ = 42, $N_C$ = 42)*

## Non-Motor Symptom Responses in Order of Decreasing Applicability (cont)

**(a) Combined scores –
DBS recipients plus carers**

**(b) Differences in scores –
(DBS recipients – carers)**

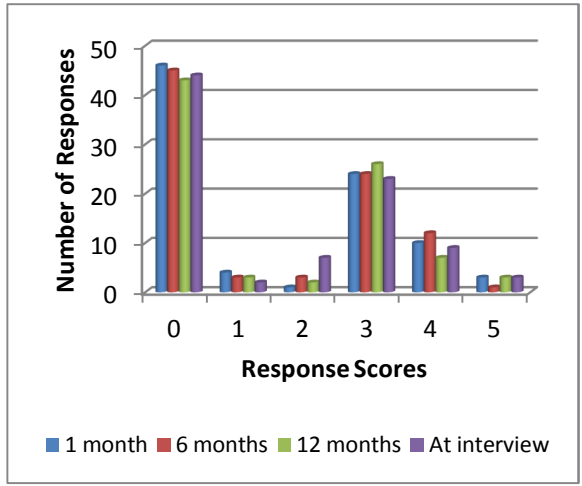
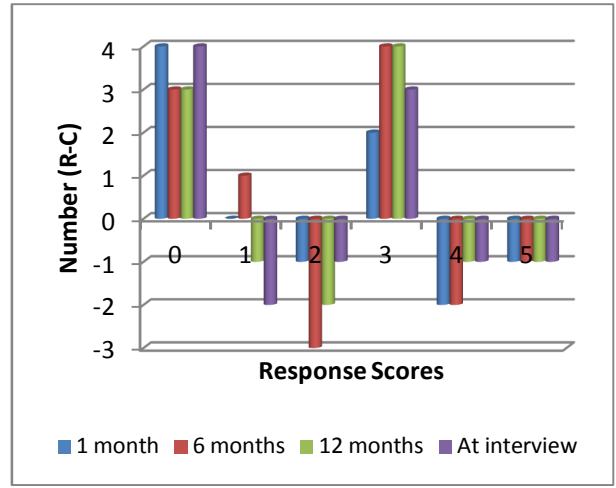

**Fig. 23 Apathy** *($N_R = 45$, $N_C = 43$)*

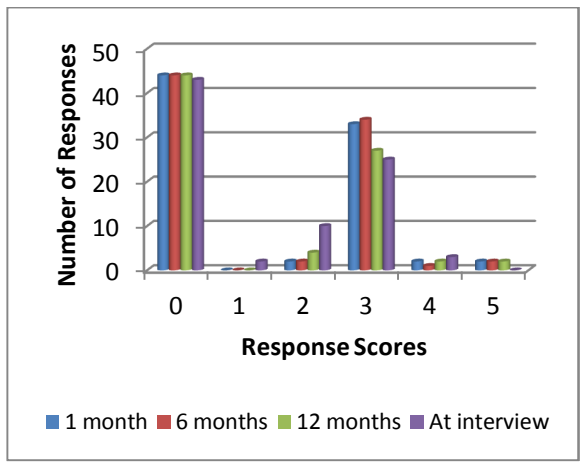
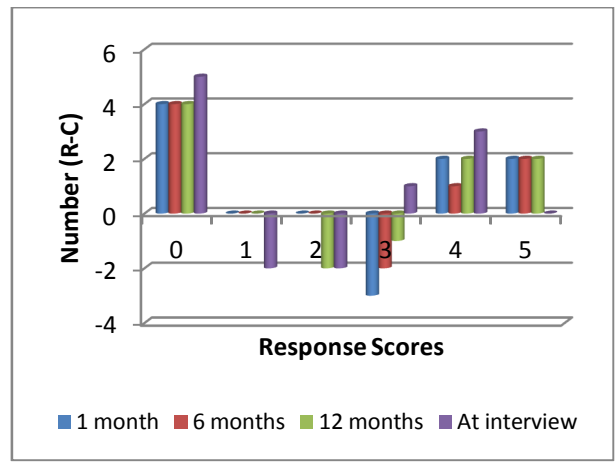

**Fig. 24 Vision** *($N_R = 44$, $N_C = 39$)*

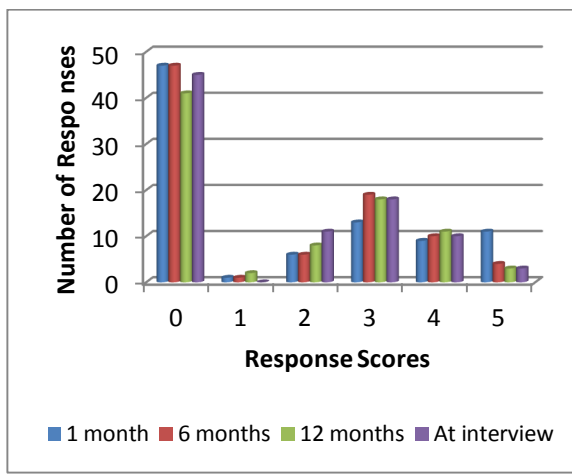
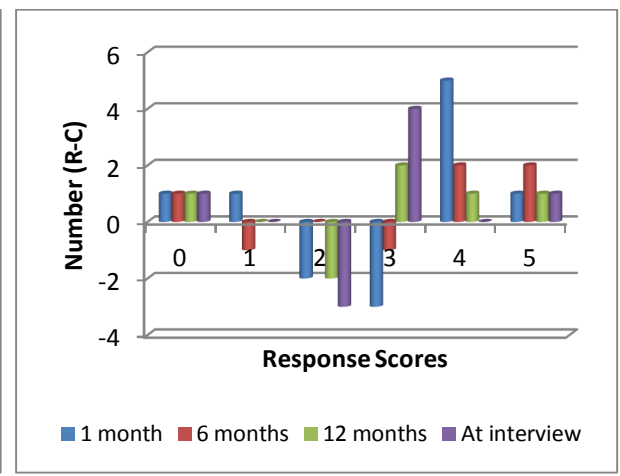

**Fig. 25 Depression** *($N_R = 45$, $N_C = 42$)*

**Non-Motor Symptom Responses in Order of Decreasing Applicability (cont)**

**(a) Combined scores –**
 DBS recipients plus carers

**(b) Differences in scores –**
 (DBS recipients – carers)

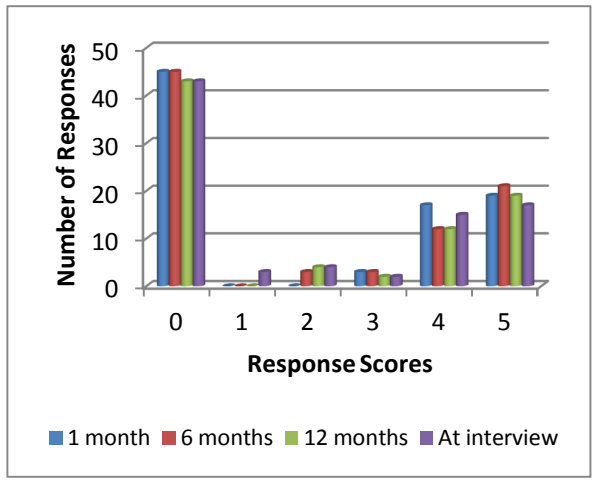
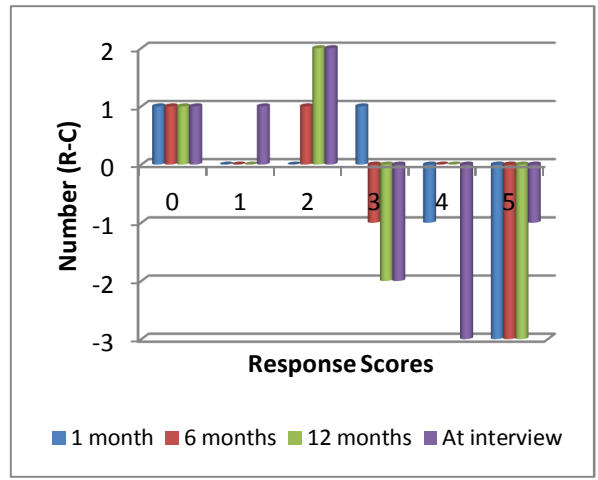

**Fig. 26 Dyskinesias** *($N_R = 41$, $N_C = 43$)*

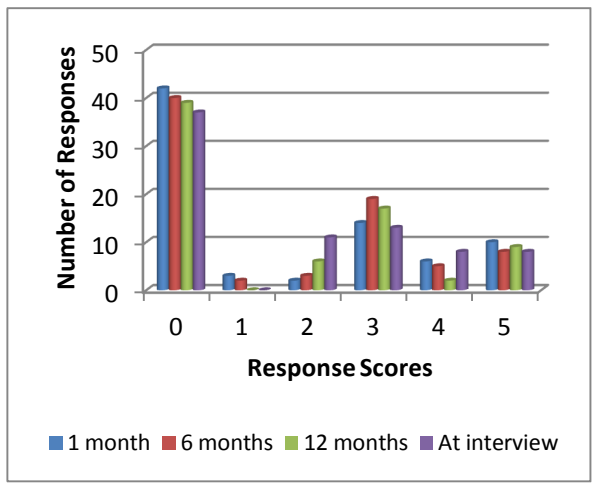
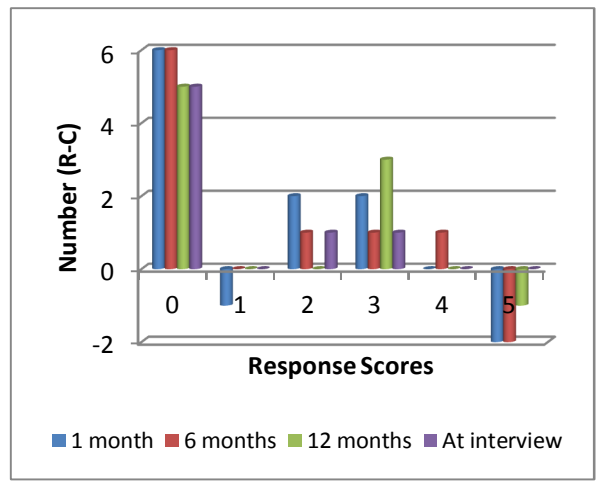

**Fig. 27 Pain** *($N_R = 42$, $N_C = 35$)*

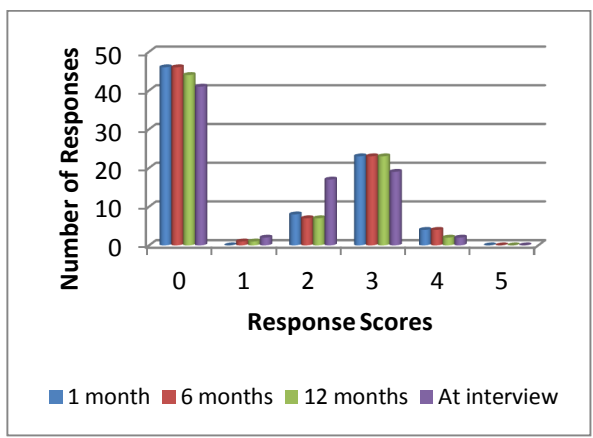
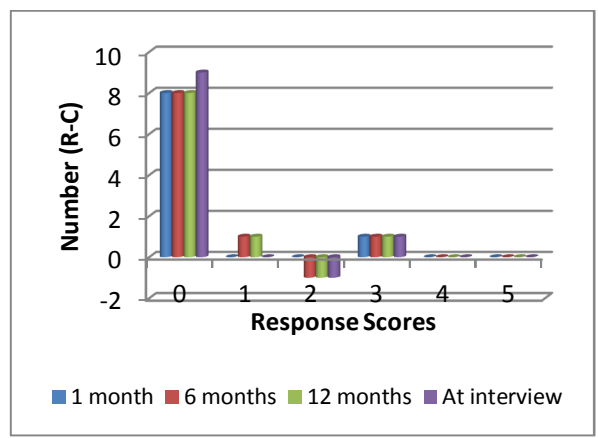

**Fig. 28 Incontinence** *($N_R = 45$, $N_C = 36$)*

**Non-Motor Symptom Responses in Order of Decreasing Applicability (cont)**

**(a) Combined scores –
DBS recipients plus carers**

**(b) Differences in scores –
(DBS recipients – carers)**

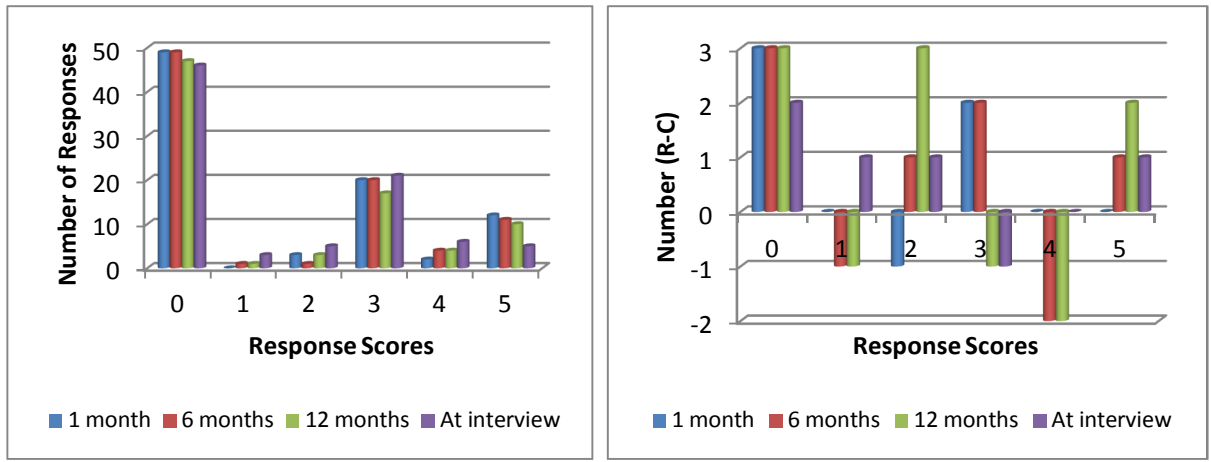

**Fig. 29 Restless leg** *($N_R$ = 45, $N_C$ = 41)*

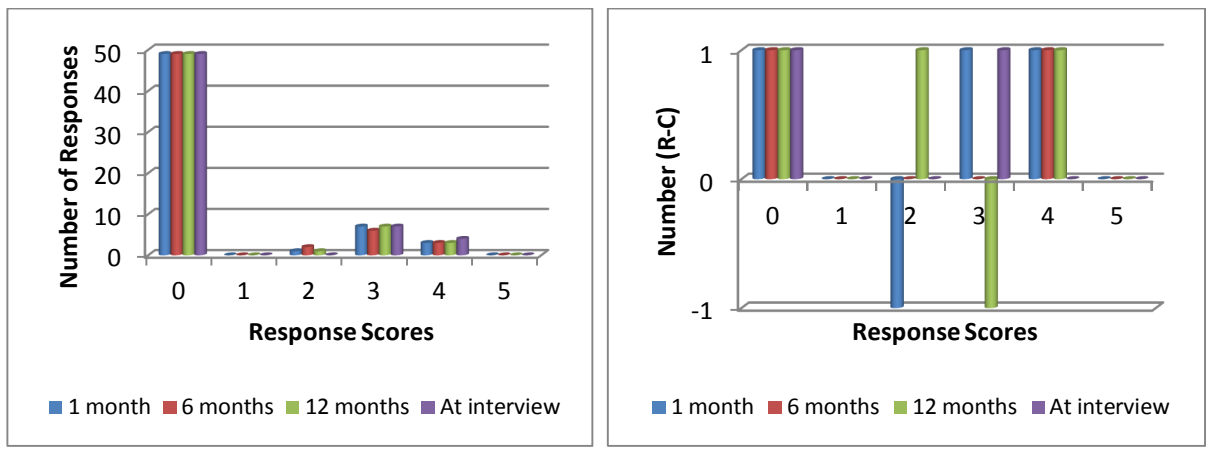

**Fig. 30 Drenching sweats (Thermoregulatory dysfunction)** *($N_R$ = 31, $N_C$ = 29)*

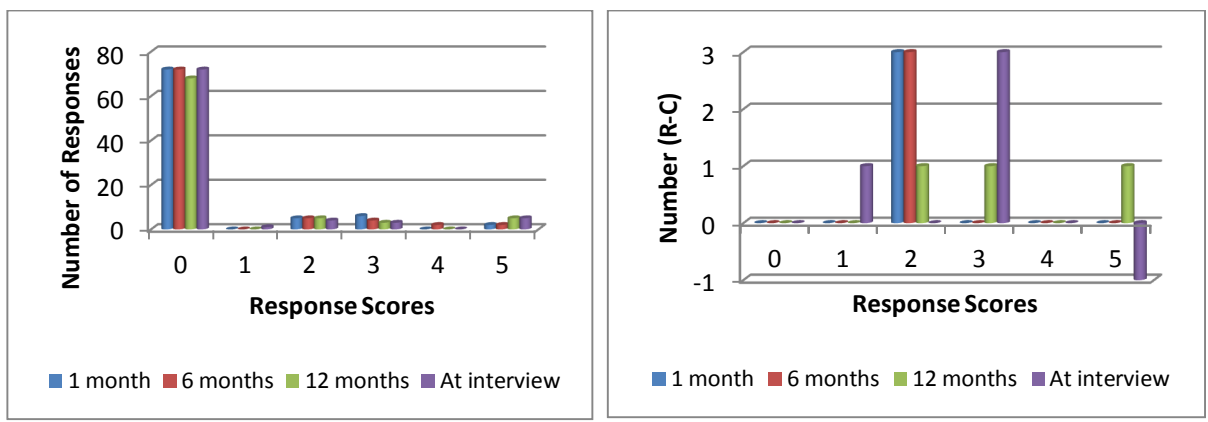

**Fig. 31 Hallucinations or delusions** *($N_R$ = 44, $N_C$ = 41)*

### 3. Quality of life changes

Two questions were included relating to quality of life – *social interaction* and *overall quality of life*. Anecdotal accounts prior to this study suggested that some with PD became more introverted and isolated as the illness progressed. These questions were included to determine what changes occurred to these aspects following DBS. The question about changes in overall quality of life was added early in the interview stage of the survey because of its frequent mention in unstructured parts of the survey. As a result formal responses to this question were only available from 74 participants.

| Table 21 – Relevance (Applicability) of Quality of Life Changes to Individual Cases | | | |
|---|---|---|---|
| Topic | Results | Reported initially as "Not Applicable" | |
| | Figure number | Participant count | Percentage |
| Overall quality of life | 32 | 0/74 | 0 |
| Social interaction | 33 | 6/90 | 7 |

| Table 22– Average Scores (1 – 5) for Quality of Life Listed in Decreasing Order of Applicability | | | | | | | | |
|---|---|---|---|---|---|---|---|---|
| Topic | DBS + 1 month | | DBS + 6 months | | DBS + 12 months | | At interview | |
| | R | C | R | C | R | C | R | C |
| Overall quality of life | 4.30 | 4.22 | 4.30 | 4.24 | 4.17 | 4.00 | 4.08 | 3.76 |
| Social interaction | 3.66 | 3.64 | 3.74 | 3.58 | 3.44 | 3.24 | 3.30 | 3.17 |

*Overall quality of life* scores were among the highest given in this survey. A high proportion gave scores of 5 (i.e. much better after DBS). Participants who had experienced a number of difficulties in the post-DBS period often still gave high scores to this feature. Details provided in Table 22 and Fig. 32 show that there is some reduction in scoring over time – an increase in the number of scores of 2 with time and a monotonic decrease in the number of scores of 5 over time, but many reported high scores over all time intervals covered in the survey. However, it should be noted that some recipients tended to give higher scores than some carers (Fig. 32b), particularly at the time of interview at which time average scores were 4.08/3.76 (R/C). These results are consistent with the results of Lewis et al. (2014) (Ref. 22).

*Social interaction* (Table 22 and Fig. 33) scores were more mixed. The most common score for the first 12 months following DBS was 3 (much the same after DBS), but a significant number also gave scores of 4 (better after DBS) or 5 (much better after DBS). But Fig. 33 shows that there was a decline in scores over the time periods covered such that, at the time of interview, the most common score was 2 (worse after DBS). For this topic also some recipients tended to give higher scores than some carers (Fig. 33b). At interview the average scores were 3.30/3.17 (R/C).

The results for *Overall quality of life* and *Social interaction* were in broad agreement with the holistic assessments of DBS in Part A (Table 3).

**Life-style Responses in Order of Decreasing Applicability**

(a) Combined scores –
DBS recipients plus carers

(b) Differences in scores –
(DBS recipients – carers)

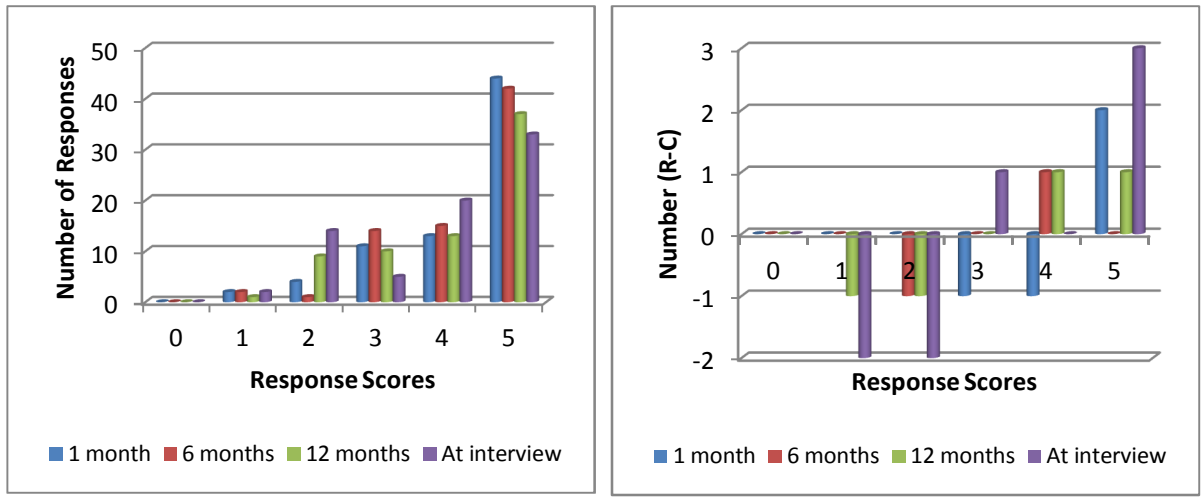

**Fig. 32 Overall Quality of Life** *($N_R$ = 37, $N_C$ = 37)*

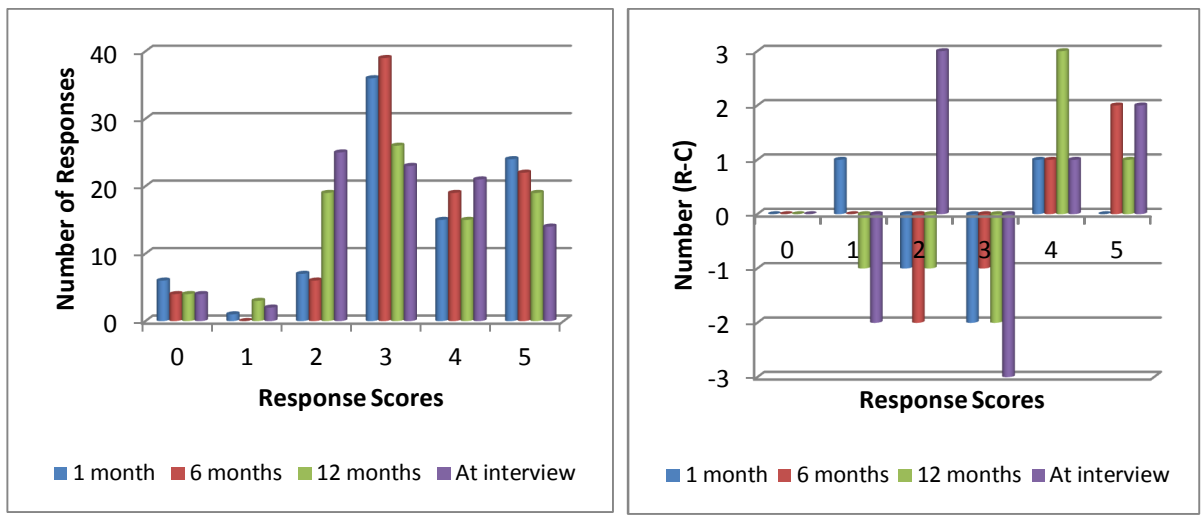

**Fig. 33 Social Interaction** *($N_R$ = 44, $N_C$ = 45)*

## 4. Summary of Part B results

Table 23 below clusters experiences with symptoms on the basis of average scores 12 months following DBS and at interview. Details of the variation between participants and with time are shown in the Figures in earlier sections of this report and in Tables 18 and 20.

| \multicolumn{5}{c}{Table 23– Summary of Perceived Changes to Symptoms Following DBS} |||||
|---|---|---|---|---|
| Average Perceived Outcome | Motor symptoms 12 months after DBS | Motor symptoms at interview (Av. 40.1 months after DBS) | Non-motor symptoms 12 months after DBS | Non-motor symptoms at interview (Av. 40.1 months after DBS) |
| Better or much better after DBS (score above 4) | Tremor | Tremor | Dyskinesias | |
| About the same or better after DBS (score range 3-4) | Difficulty Walking, Slowness or lack of movement, Facial expression, Difficulty standing from a chair, Difficulty turning in bed, Stiffness and rigidity, Freezing | Slowness or lack of movement, Facial expression, Difficulty turning in bed, Stiffness and rigidity | Sleep quality, Fatigue, Sense of Smell, Mood and behaviour, Constipation, Anxiety, Severity of on/off periods, Dystonia, Apathy, Vision, Depression, Pain, Restless leg, Thermoregulatory dysfunction, Hallucinations or delusions | Dyskinesias, Sleep quality, Sense of Smell, Mood and behaviour, Anxiety, Severity of on/off periods, Dystonia, Apathy, Depression, Pain, Restless leg, Thermoregulatory dysfunction, Hallucinations or delusions |
| Worse after DBS (score less than 3) | Handwriting, Speech, Postural stability, Swallowing | Handwriting, Difficulty walking, Speech, Postural stability, Difficulty standing from a chair, Swallowing, Freezing | Cognitive function, memory and reasoning ability, Incontinence | Fatigue, Cognitive function, memory and reasoning ability, Constipation, Vision, Incontinence |

# CONCLUSIONS

This study has provided a new perspective on the outcomes from DBS for Parkinson's disease by sampling the experiences of both recipients of DBS and also the carers. Information was collected for 52 cases involving 91 participants (46 recipients and 45 carers). The results provide new information about the frequency of different symptoms in Parkinson's disease and the perceived impact of these symptoms following the DBS procedure, and how these impacts change with time.

DBS is an exacting surgical treatment for PD. In 27% of the cases covered in this survey, further DBS related surgery was necessary to achieve or restore optimal benefits from the procedure. In a number of cases the circumstances leading to further surgery represented a very stressful episode for the DBS recipient and the carer, but most were ultimately pleased with the outcomes following this additional intervention.

There was a high degree of consistency between Part A and Part B of the survey. The open- ended section of the survey (Part A) revealed many features of importance to DBS recipients and carers, ranging from changes in PD management, changes in symptoms and impact on their lifestyle. Their comments were in broad agreement with the scaling of symptom changes in Part B of the survey.

On average, the outcomes from DBS for PD revealed in this survey were in broad agreement with previously published results showing improvements across a wide range of symptoms. However in this study the outcomes varied significantly between different DBS recipients and over time.

Significant improvements in quality of life, reduced or eliminated tremor and dyskinesia were reported by many. These improvements were often sustained over long periods.

A wide range of motor and non-motor symptoms were reported to be at least the same or better after DBS, but in some cases there was considerable variability in experiences between participants and over time. The improvements in some symptoms following DBS were sustained, but others declined with time, some noticeably even within 12 months of the procedure.

Some symptoms appeared worse on average after DBS – these included postural stability, speech and cognitive function.

While there was often broad agreement between the assessments of outcomes from DBS recipients on the one hand, and carers on the other, there were noticeable differences for some symptoms and outcomes at different times following DBS. When this happened, the carer tended to give a less positive assessment of the improvement following DBS. Examples included sleep quality, fatigue, cognitive function, memory and reasoning ability, mood and behaviour, anxiety, vision, depression and quality of life.

<p style="text-align:center">\*\*\*</p>

# APPENDIX 1 – SURVEY INSTRUMENT

**Preamble**

People contemplating Deep Brain Stimulation (DBS) for Parkinson's disease often ask us questions like "what can I expect to experience after the procedure?". We are conducting a survey of those who have had experience with DBS to help us provide more widely based answers to such enquiries. The results of our survey will be written into a report which, together with other information, will be made available to those interested in the range of outcomes reported by the participants in this survey.

What follows DBS not only affects those who have had the procedure but also those that care for them, family members and friends. The treating neurologist is the primary reference point for information, but access to the neurologist is restricted to periodic visits and not always available to carers, family and friends. Importantly, those who have had the procedure are in a unique position to provide a holistic commentary on their journey with PD following DBS, particularly how they have changed over time and how in turn those changes have affected family, friends and carers.

PD is a highly individualistic disease – every person with PD has their own story. This is true with or without DBS. An important contribution to the preparation for life after DBS is an understanding of the mosaic of previous experiences of those who have gone before – as many as possible to provide a realistic appreciation of the commonly experienced phases of recovery, changes in the management of PD and the impact on loved ones.

As someone who has had DBS we are approaching you to see if you are willing to help us with this important task by sharing with us your experiences. Any information provided will be treated in confidence. Any written accounts of the survey will have personal information removed to preserve the anonymity of participants unless they have previously provided written consent for them to be identified. There is no compulsion about any aspect of this survey – potential participants are free to decline to participate or to withdraw at any time without the need to give any reason.

Participants sought for this survey include those who have had DBS for PD, and also where possible their carers and family so as to gain the holistic response that we seek. In the first instance we are sending you this letter so that you can see the questions we wish to ask. A short time later we will phone you to determine your willingness to participate, and if you are, to take you through the questions of the survey. We would like to repeat the process with the carer so that both viewpoints are captured.

I sincerely hope that you can help us in this way.

## Personal background

Date of interview:

|   |                                  | **DBS for PD** |
|---|----------------------------------|----------------|
| 1 | Name (optional)                  |                |
| 2 | Age                              |                |
| 3 | Gender                           |                |
| 4 | Date diagnosed PD                |                |
| 5 | Previous illness to PD           |                |
| 6 | Date DBS implanted               |                |
| 7 | DBS site?                        |                |
| 8 | Other illnesses in addition to PD|                |
| 9 | PD medication before DBS         |                |

**Procedure History**

1. Have there been further periods in hospital related to DBS **requiring surgery** after the initial implant procedure? (e.g. battery replacement, correction of adverse effects etc). If so please provide dates and details.
2. Have there been further periods in hospital related to DBS but **not requiring further surgery** after the initial implant procedure? (e.g. physiotherapy, speech therapy etc). If so please provide dates and details.

**PD symptoms following DBS**

We would like to ask you two series of questions – Part A, your description of the changes you noticed and Part B, a quantitatively scaled comparison of changes to symptoms after DBS. (There may be some overlap between parts A and B)

PD symptoms change over time, with or without DBS. In this section we are asking you to rate your **perception** of symptom severity compared with your recollections of what they were **before** DBS.

## Part A

Briefly describe the most important changes you noticed over the period indicated in the table.

| Period | Most important changes (PwP) |
|---|---|
| DBS + 1 month | Medication?<br>Frequency of DBS setting changes and your experiences following the change?<br>Other? |
| DBS + 6 months | Medication?<br>Frequency of DBS setting changes and your experiences following the change?<br>Other? |
| DBS + 12 months | Medication?<br>Frequency of DBS setting changes and your experiences following the change?<br>Other? |
| DBS to now | Medication?<br>Frequency of DBS setting changes and your experiences following the change?<br>Other? |

| Period | Most important changes (Carer) |
|---|---|
| DBS + 1 month |  |
| DBS + 6 months |  |
| DBS + 12 months |  |
| DBS to now |  |

| **General Comments on DBS Experience (PwP and Carer)** |
|---|
|  |

## Part B

In this section we would like you to compare your perception of your symptoms before and after DBS at different times relevant to your situation. Where appropriate use the scale:

5 = much better after DBS   2 = worse after DBS

4 = better after DBS   1 = significantly worse after DBS

3 = much the same after DBS   0 = not applicable

| | | PwP | | | | Carer | | | |
| --- | --- | --- | --- | --- | --- | --- | --- | --- | --- |
| | | Time since DBS (months) | | | | Time since DBS (months) | | | |
| | Common Symptoms | 1 | 6 | 12 | Now | 1 | 6 | 12 | Now |
| 1 | Tremor? | | | | | | | | |
| 2 | Stiffness / rigidity? | | | | | | | | |
| 3 | Slowness or lack of movement? | | | | | | | | |
| 4 | Postural instability – unsteadiness, falls? | | | | | | | | |
| 5 | Difficulty walking? | | | | | | | | |
| 6 | Difficulty standing from a chair? | | | | | | | | |
| 7 | Freezing? | | | | | | | | |
| 8 | Speech? | | | | | | | | |
| 9 | Handwriting? | | | | | | | | |
| 10 | Swallowing? | | | | | | | | |
| 11 | Constipation? | | | | | | | | |
| 12 | Incontinence? | | | | | | | | |
| 13 | Sleep quality (disruption/dreams etc)? | | | | | | | | |
| 14 | Turning over in bed? | | | | | | | | |
| 15 | Restless leg? | | | | | | | | |
| 16 | Pain? | | | | | | | | |
| 17 | Fatigue? | | | | | | | | |
| 18 | Apathy? | | | | | | | | |
| 19 | Anxiety? | | | | | | | | |
| 20 | Depression? | | | | | | | | |
| 21 | Hallucinations or delusions? | | | | | | | | |
| 22 | Sense of smell? | | | | | | | | |
| 23 | Vision? | | | | | | | | |
| 24 | Cognitive function, memory and reasoning ability? | | | | | | | | |
| 25 | Facial expression? | | | | | | | | |
| 26 | Dyskinesias? | | | | | | | | |
| 27 | Severity of on/off periods? | | | | | | | | |
| 28 | Dystonia? (sustained muscle contractions) | | | | | | | | |
| 29 | Changes in mood and behaviour? | | | | | | | | |
| 30 | Drenching sweats (thermoregulatory dysfunction)? | | | | | | | | |
| 31 | Social interaction? | | | | | | | | |
| 32 | Overall quality of life? | | | | | | | | |
| 33 | Any other symptom, not listed above, that is important to your case (please specify)? | | | | | | | | |